\documentclass[aps,prc,preprint,groupedaddress,showpacs]{revtex4}
\usepackage{graphicx}

\begin{document}


\title{Hartree-Fock approach to nuclear matter and finite nuclei\\
with M3Y-type nucleon-nucleon interactions}


\author{H. Nakada}
\email[E-mail:\,\,]{nakada@faculty.chiba-u.jp}
\affiliation{Department of Physics, Faculty of Science, Chiba University\\
Yayoi-cho 1-33, Inage, Chiba 263-8522, Japan}


\date{\today}

\begin{abstract}
By introducing a density-dependent contact term,
M3Y-type interactions applicable to the Hartree-Fock calculations
are developed.
In order to view basic characters of the interactions,
we carry out calculations on the uniform nuclear matter
as well as on several doubly magic nuclei.
It is shown that a parameter-set called M3Y-P2
describes various properties
similarly well to the Skyrme SLy5 and/or the Gogny D1S interactions.
A remarkable difference from the SLy5 and the D1S interactions
is found in the spin-isospin properties in the nuclear matter,
to which the one-pion-exchange potential gives
a significant contribution.
Affecting the single-particle energies,
this difference may play a certain role in the new magic numbers
in unstable nuclei.
\end{abstract}


\maketitle



\section{Introduction\label{sec:intro}}

Various models for nuclear structure have been developed
in order to study low energy phenomena of the atomic nuclei.
Whereas straightforward application of the bare $NN$ interaction
is yet limited only to light nuclei~\cite{ref:micro},
the nuclear structure seems to be well described
by relatively simple effective interactions at low energies.
Although the effective interactions may depend on the models,
there should be basic characters in the effective interactions
for the low energy phenomena,
irrespective to the models.
On the other hand, since the invention of the secondary beam technology,
experimental data on the unstable nuclei
have disclosed new aspects of the nuclear structure.
A remarkable example is the dependence of magic numbers
on the neutron excess~\cite{ref:magic}.
In regard to the new magic numbers discovered near the neutron drip line,
a question has been raised
on a character of the effective interactions
relating to the spin-isospin flip mode~\cite{ref:Vst}.

Mean-field theories have successfully been applied
to the nuclear structure problems,
in particular for stable nuclei.
They are also useful to investigate basic characters
of the effective interactions.
However, not many effective interactions have been explored
for the nuclear mean-field calculations so far.
The Skyrme interaction~\cite{ref:VB72} has been popular
in the Hartree-Fock (HF) calculations,
since the zero-range form is easy to be handled.
Among a limited number of finite-range interactions,
the Gogny interaction~\cite{ref:Gogny} is widely applied
to the mean-field calculations,
in which the Gaussian form is assumed for the central force.
The parameter-sets both of the Skyrme and the Gogny interactions
have been adjusted mainly to the data on the nuclei
around the $\beta$-stability.
It is not obvious whether the available parameter-sets
of these interactions account for the new magic numbers properly.

In order to exploit effective interactions
applicable also to unstable nuclei,
guide from microscopic theories will be important.
Brueckner's $G$-matrix has been a significant clue to studies
in this course.
Although microscopic approaches using the $G$-matrix
have not yet been successful
in reproducing the saturation properties,
notable progress has been made recently.
In the shell model approaches,
microscopic effective interactions have been shown
to reproduce observed levels remarkably well~\cite{ref:shell-G}.
It should be noted, however, that the shell model interactions are
usually specific to mass regions,
and their global characters have not been discussed in detail,
despite several exceptions~\cite{ref:ST}.
The so-called Michigan 3-range Yukawa (M3Y) interaction~\cite{ref:M3Y}
has been derived from the bare $NN$ interaction,
by fitting the Yukawa functions to the $G$-matrix.
Represented by the sum of the Yukawa functions,
the M3Y-type interactions will be tractable
in various models.
It has been shown that the M3Y interaction gives
matrix elements similar to reliable shell model interactions~\cite{ref:USD}.
Moreover, with a certain modification,
M3Y-type interactions have successfully been applied
to nuclear reactions~\cite{ref:reac}.
By using a recently developed algorithm~\cite{ref:NS02},
a class of the M3Y-type interactions can be applied
also to the mean-field calculations.
Under such circumstances,
it will be of interest to explore M3Y-type interactions
and to investigate their characters in the mean-field framework.
In this article, we shall develop M3Y-type interactions
and investigate their characters via the HF calculations.

\section{Modification of M3Y interaction\label{sec:M3Y}}

Nuclear effective Hamiltonian consists
of the kinetic energy and the effective interaction,
\begin{equation}
H = K + V\,;\quad K = \sum_i \frac{\mathbf{p}_i^2}{2M}\,,\quad
V = \sum_{i<j} v_{ij}\,.
\end{equation}
Here $i$ and $j$ are the indices of individual nucleons.
It will be natural to assume the effective interaction $v_{ij}$
to be translationally invariant,
except for the density dependence mentioned below.
We consider the effective interaction having the following form,
\begin{eqnarray} v_{12} &=& v_{12}^{(\mathrm{C})}
 + v_{12}^{(\mathrm{LS})} + v_{12}^{(\mathrm{TN})}
 + v_{12}^{(\mathrm{DD})}\,;\nonumber\\
v_{12}^{(\mathrm{C})} &=& \sum_n (t_n^{(\mathrm{SE})} P_\mathrm{SE}
+ t_n^{(\mathrm{TE})} P_\mathrm{TE} + t_n^{(\mathrm{SO})} P_\mathrm{SO}
+ t_n^{(\mathrm{TO})} P_\mathrm{TO})
 f_n^{(\mathrm{C})} (r_{12})\,,\nonumber\\
v_{12}^{(\mathrm{LS})} &=& \sum_n (t_n^{(\mathrm{LSE})} P_\mathrm{TE}
 + t_n^{(\mathrm{LSO})} P_\mathrm{TO})
 f_n^{(\mathrm{LS})} (r_{12})\,\mathbf{L}_{12}\cdot
(\mathbf{s}_1+\mathbf{s}_2)\,,\nonumber\\
v_{12}^{(\mathrm{TN})} &=& \sum_n (t_n^{(\mathrm{TNE})} P_\mathrm{TE}
 + t_n^{(\mathrm{TNO})} P_\mathrm{TO})
 f_n^{(\mathrm{TN})} (r_{12})\, r_{12}^2 S_{12}\,,\nonumber\\
v_{12}^{(\mathrm{DD})} &=& t^{(\mathrm{DD})}
 (1 + x^{(\mathrm{DD})} P_\sigma)
 [\rho(\mathbf{r}_1)]^\alpha \delta(\mathbf{r}_{12})\,.
\label{eq:effint}\end{eqnarray}
The relative coordinate is denoted
by $\mathbf{r}_{12}= \mathbf{r}_1 - \mathbf{r}_2$
and $r_{12}=|\mathbf{r}_{12}|$.
Correspondingly, the relative momentum is defined by
$\mathbf{p}_{12}= (\mathbf{p}_1 - \mathbf{p}_2)/2$.
$\mathbf{L}_{12}$ is the relative orbital angular momentum,
\begin{equation} \mathbf{L}_{12}= \mathbf{r}_{12}
\times \mathbf{p}_{12}\,,
\label{eq:L12}\end{equation}
$\mathbf{s}_1$, $\mathbf{s}_2$ are the nucleon spin operators,
and $S_{12}$ is the tensor operator,
\begin{equation} S_{12}= 4\,[
3(\mathbf{s}_1\cdot\hat{\mathbf{r}}_{12})
(\mathbf{s}_2\cdot\hat{\mathbf{r}}_{12})
- \mathbf{s}_1\cdot\mathbf{s}_2 ]\,.
\label{eq:tensor}\end{equation}
$f_n(r_{12})$ represents an appropriate function of $r_{12}$,
the subscript $n$ corresponds to the parameter attached to the function
(\textit{e.g.} the range of the interaction),
and $t_n$ is the coefficient.
Examples of $f_n(r_{12})$ are
the delta, the Gauss and the Yukawa functions.
$P_\sigma$ ($P_\tau$) denotes the spin (isospin) exchange operator,
while $P_\mathrm{SE}$, $P_\mathrm{TE}$, $P_\mathrm{SO}$
and $P_\mathrm{TO}$ are the projection operators
on the singlet-even (SE), triplet-even (TE), singlet-odd (SO)
and triplet-odd (TO) two-particle states, respectively,
which are defined by
\begin{eqnarray}
P_\mathrm{SE} = \frac{1-P_\sigma}{2}\,\frac{1+P_\tau}{2}\,,
\quad P_\mathrm{TE} = \frac{1+P_\sigma}{2}\,\frac{1-P_\tau}{2}\,,
\nonumber\\
P_\mathrm{SO} = \frac{1-P_\sigma}{2}\,\frac{1-P_\tau}{2}\,.
\quad P_\mathrm{TO} = \frac{1+P_\sigma}{2}\,\frac{1+P_\tau}{2}\,.
\label{eq:proj_T}\end{eqnarray}
The nucleon density is denoted by $\rho(\mathbf{r})$.
The original M3Y interaction is represented
in the form of Eq.~(\ref{eq:effint}),
with $f_n(r_{12})=e^{-\mu_n r_{12}}/\mu_n r_{12}$
and $v_{12}^{(\mathrm{DD})}=0$.
As discussed in Ref.~\cite{ref:NS02},
the Skyrme and the Gogny interactions are obtained
by setting $f_n(r_{12})$ appropriately,
except for some parameter-sets of the Skyrme interaction
in which certain terms are expressed
only in the density-functional form.

The saturation of density and energy is a basic property of nuclei.
In developing effective interactions adaptable for many nuclei,
it is required to reproduce the saturation property.
However, the non-relativistic $G$-matrix fails to reproduce
the saturation at the right density and energy.
Therefore, it will not be appropriate
to use the $G$-matrix for HF calculations
without any modification,
although several HF approaches
using interactions derived from the $G$-matrix were tried
in earlier studies~\cite{ref:Neg70}.
The M3Y interaction was obtained so that
the $G$-matrix at a certain density could be reproduced
by a sum of the Yukawa functions.
The M3Y interaction gives no saturation point
within the HF theory,
unless density-dependence is taken into account explicitly.
Khoa \textit{et al.} applied the M3Y interaction
to nuclear reactions in the folding model,
by making the coupling constants dependent on densities~\cite{ref:reac}.
In their approach the exchange terms are treated approximately.
However, the exchange terms may contribute quite significantly
to the nuclear structure.
We here keep the coupling constants in $v_{12}^{(\mathrm{C})}$
independent of density,
while introduce a density-dependent contact interaction
($v_{12}^{(\mathrm{DD})}$ in Eq.~(\ref{eq:effint})),
as in the Skyrme and the Gogny interactions.
We can then treat the exchange (\textit{i.e.} the Fock) terms exactly
with the currently available computers.
It should be mentioned that there has been an interesting attempt
to approximate the exchange terms
of the interaction in the density-matrix expansion~\cite{ref:HL98},
although the accuracy of the density-matrix expansion
should be checked carefully.

We start from the Paris-potential version
of the M3Y interaction~\cite{ref:M3Y-P}.
This original parameter-set with no density-dependence
is hereafter called `M3Y-P0'.
We shall modify this interaction
so as to reproduce the saturation properties.
In the isotropic uniform nuclear matter,
matrix elements of $v_{12}^{(\mathrm{LS})}$ and $v_{12}^{(\mathrm{TN})}$
between the HF states vanish.
Therefore $v_{12}^{(\mathrm{C})}+v_{12}^{(\mathrm{DD})}$ determines
the bulk properties such as the saturation.
The range parameters for the Yukawa functions
$f^{(\mathrm{C})}_n(r_{12})=e^{-\mu_n r_{12}}
/\mu_n r_{12}$ in $v_{12}^{(\mathrm{C})}$
are $\mu_1^{-1}=0.25$, $\mu_2^{-1}=0.4$
and $\mu_3^{-1}=1.414\,\mathrm{fm}$ in the M3Y interaction,
which correspond to the Compton wave-lengths of mesons
with masses of about 790, 490 and 140\,MeV, respectively.
We do not change these parameters.
For the longest-range part ($n=3$),
the coupling constants $t_3^{(\mathrm{SE})}$, $t_3^{(\mathrm{TE})}$,
$t_3^{(\mathrm{SO})}$ and $t_3^{(\mathrm{TO})}$ are fixed
to be those of the one-pion-exchange potential (OPEP),
as in M3Y-P0.
The interaction $v_{12}^{(\mathrm{DD})}$ in Eq.~(\ref{eq:effint})
acts only on the SE and TE channels,
\begin{equation} v_{12}^{(\mathrm{DD})}
= t^{(\mathrm{DD})} (1-x^{(\mathrm{DD})})
 \delta(\mathbf{r}_{12}) P_\mathrm{SE}
+ t^{(\mathrm{DD})} (1+x^{(\mathrm{DD})})
 \delta(\mathbf{r}_{12}) P_\mathrm{TE}\,.
\label{eq:DDchannel}\end{equation}
Microscopic investigations have shown
that the density-dependence of the TE part
is primarily responsible for the saturation~\cite{ref:Bet71},
as a higher-order effect of the tensor force.
While the interaction in the SE channel is attractive at low densities,
it also has certain density-dependence
originating in the strong short-range repulsion.
Thus, a possible way of modifying the M3Y interaction
may be to replace a fraction of the repulsion in the SE and TE channels
by $v_{12}^{(\mathrm{DD})}$.

In addition to the saturation properties
which are relevant to the central force,
the LS splitting is significant
in describing the shell structure of nuclei.
While true origin of the LS splitting
is not yet obvious~\cite{ref:LS},
LS splittings obtained from HF calculations
with the $G$-matrix interaction are too small,
in comparison with the observed ones.
From the HF calculations for finite nuclei,
we find that $v_{12}^{(\mathrm{LS})}$ should be about twice as strong as
that of M3Y-P0 to reproduce the observed LS splittings.
The tensor force influences the ordering
of the single-particle (s.p.) orbits.
To reproduce the observed ordering,
$v_{12}^{(\mathrm{TN})}$ should be smaller than that of M3Y-P0.
We here introduce an overall enhancement factor to $v_{12}^\mathrm{(LS)}$
and an overall reduction factor to $v_{12}^\mathrm{(TN)}$,
as will be shown in Section~\ref{sec:DMprop}.

In this paper we shall use two parameter-sets
for modified M3Y interaction, `M3Y-P1' and `M3Y-P2',
in order to show sensitivity to the parameters for some results.
In M3Y-P1, we replace the shortest-range ($n=1$) repulsive part
of $v_{12}^\mathrm{(C)}$ by $v_{12}^{(\mathrm{DD})}$
in a simple manner.
We reduce both $t_1^{(\mathrm{SE})}$
and $t_1^{(\mathrm{TE})}$ by a single factor,
keeping the SE/TE ratio in $v_{12}^{(\mathrm{DD})}$
to be equal to $t_1^{(\mathrm{SE})}/t_1^{(\mathrm{TE})}$ in M3Y-P0,
by imposing
\begin{equation} x^{(\mathrm{DD})}
 = \frac{t_1^{(\mathrm{TE})}-t_1^{(\mathrm{SE})}}
{t_1^{(\mathrm{TE})}+t_1^{(\mathrm{SE})}}\,.
\label{eq:x3-a}\end{equation}
The reduction factor and $t^{(\mathrm{DD})}$ are determined
so as for the saturation density and energy in the nuclear matter
to be typical values,
as presented in the subsequent section.
Characters of M3Y-P1 will be investigated in the nuclear matter.
Although this modification is too simple
to reproduce properties of finite nuclei,
the M3Y-P1 set will be useful to clarify
what characters arise from the original M3Y interaction,
relatively insensitive to the phenomenological modification.
In the M3Y-P2 set, all $t_n$ parameters
belonging to the $n=1$ and $2$ channels in $v_{12}^{(\mathrm{C})}$
are shifted from those of M3Y-P0.
Although we have three ranges in $v_{12}^{(\mathrm{C})}$,
the number of adjustable parameters is no greater than
in the Gogny interaction, since we fix the OPEP part.
We fit those parameters,
together with the enhancement factor for $v_{12}^{(\mathrm{LS})}$
and the reduction factor for $v_{12}^{(\mathrm{TN})}$,
to the binding energies of several doubly magic nuclei.
The resultant values of the parameters will be shown later.

\section{Properties of nuclear matter
at and around saturation point\label{sec:NMprop}}

Basic characters of nuclear effective interactions
can be discussed via properties of the infinite nuclear matter;
in particular, properties at and around the saturation point.
In this section we investigate characters of the M3Y-type interactions
via the nuclear matter properties within the HF theory.
In comparison, we also discuss those of the Skyrme
and the Gogny interactions.
We use the D1S parameter-set~\cite{ref:D1S} for the Gogny interaction.
In most of the Skyrme HF approaches,
the LS currents arising from the momentum-dependence of the central force
are ignored,
and the parameters are adjusted without their contribution.
Although this treatment occasionally improves
some characters of the interactions,
in this paper we would focus on characters of the two-body interactions,
rather than those of density functionals.
For this reason we adopt the SLy5 set~\cite{ref:SLy},
which is devised for calculations including the LS currents.

In the HF theory of the nuclear matter,
the s.p. wave-functions can be taken to be the plane wave,
\begin{equation} \varphi_{\mathbf{k}\sigma\tau}(\mathbf{r})
= \frac{1}{\sqrt\Omega}\,e^{i\mathbf{k}\cdot\mathbf{r}}\,
\chi_\sigma \chi_\tau\,.
\label{eq:NM-spwf}\end{equation}
Here $\chi_\sigma$ ($\chi_\tau$) denotes
the spin (isospin) wave-function,
and $\Omega$ indicates the volume of the system,
for which we will take the $\Omega\rightarrow\infty$ limit afterward.
The s.p. energy for this state is defined as
\begin{equation}
 \epsilon(\mathbf{k}\sigma\tau) =
  \frac{\mathbf{k}^2}{2M}
 + \frac{\Omega}{(2\pi)^3} \sum_{\sigma_2\tau_2}
  \int_{k_2\leq k_{\mathrm{F}\tau_2\sigma_2}}d^3k_2
  \langle\mathbf{k}\sigma\tau,\mathbf{k}_2\sigma_2\tau_2
  |v_{12}|\mathbf{k}\sigma\tau,\mathbf{k}_2\sigma_2\tau_2\rangle\,.
\label{eq:NM-spe}\end{equation}
Energy of the nuclear matter is expressed
by a function of densities
depending on the spin and the isospin,
$\rho_{\tau\sigma}$ ($\tau=p,n$; $\sigma=\uparrow,\downarrow$).
The density variables can be converted to the total density
$\rho={\displaystyle\sum_{\sigma\tau}} \rho_{\tau\sigma}$,
and the spin- and isospin-asymmetry parameters
\begin{eqnarray}
 \eta_s &=& \frac{{\displaystyle\sum_{\sigma\tau}}
  \sigma\rho_{\tau\sigma}}{\rho}
  ~=~ \frac{\rho_{p\uparrow}-\rho_{p\downarrow}+\rho_{n\uparrow}
  -\rho_{n\downarrow}}{\rho}\,, \nonumber\\
 \eta_t &=& \frac{{\displaystyle\sum_{\sigma\tau}}
  \tau\rho_{\tau\sigma}}{\rho}
  ~=~ \frac{\rho_{p\uparrow}+\rho_{p\downarrow}-\rho_{n\uparrow}
  -\rho_{n\downarrow}}{\rho}\,, \nonumber\\
 \eta_{st} &=& \frac{{\displaystyle\sum_{\sigma\tau}}
  \sigma\tau\rho_{\tau\sigma}}{\rho}
  ~=~ \frac{\rho_{p\uparrow}-\rho_{p\downarrow}-\rho_{n\uparrow}
  +\rho_{n\downarrow}}{\rho}\,,
\end{eqnarray}
where $\sigma$ ($\tau$) in the summation takes $\pm 1$,
corresponding to $\sigma=\uparrow,\downarrow$ ($\tau=p,n$).
By assuming that the s.p. states are occupied
up to the Fermi momentum,
the density is related to the Fermi momentum for each spin and isospin,
\begin{equation}
 \rho_{\tau\sigma} = \frac{1}{6\pi^2} k_{\mathrm{F}\tau\sigma}^3\,.
\end{equation}
The total energy of nuclear matter is given by
\begin{eqnarray}
 E &=& \frac{\Omega}{(2\pi)^3} \sum_{\sigma_1\tau_1}
  \int_{k_1\leq k_{\mathrm{F}\tau_1\sigma_1}}d^3k_1
  \frac{\mathbf{k}_1^2}{2M} \nonumber\\
 && + \frac{\Omega^2}{2(2\pi)^6} \sum_{\sigma_1\sigma_2\tau_1\tau_2}
  \int_{k_1\leq k_{\mathrm{F}\tau_1\sigma_1}}d^3k_1
  \int_{k_2\leq k_{\mathrm{F}\tau_2\sigma_2}}d^3k_2
  \langle\mathbf{k}_1\sigma_1\tau_1,\mathbf{k}_2\sigma_2\tau_2
  |v_{12}|\mathbf{k}_1\sigma_1\tau_1,\mathbf{k}_2\sigma_2\tau_2\rangle
 \,. \nonumber\\
\label{eq:NME}\end{eqnarray}
As already pointed out,
only $v_{12}^{(\mathrm{C})}+v_{12}^{(\mathrm{DD})}$
contributes to the energy of the isotropic nuclear matter.
In Appendix~\ref{app:NME}, several formulae
on the HF energy of the nuclear matter
are derived for interactions expressed in the form of Eq.~(\ref{eq:effint}),
with general and typical $f^{(\mathrm{C})}_n(r_{12})$.
As well as for the M3Y-type interactions,
the nuclear matter energies are calculated
for the Skyrme and the Gogny interactions
by using these formulae.

\begin{figure}
\includegraphics[height=10cm]{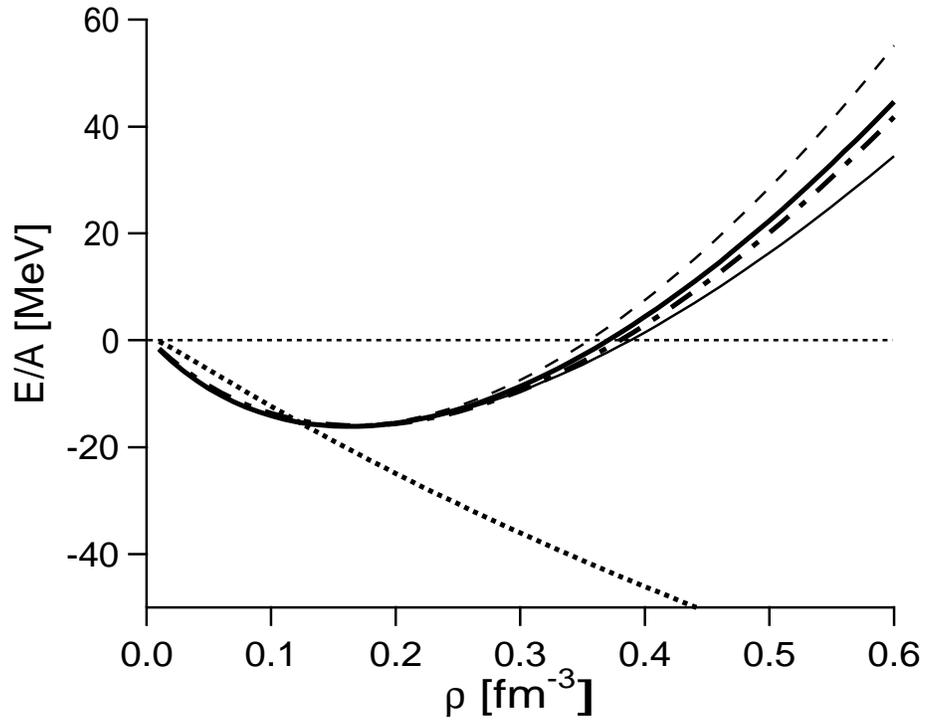}
\vspace{2mm}
\caption{Energies per nucleon $\mathcal{E}=E/A$
in the symmetric nuclear matter for several effective interactions.
The thick dotted, dot-dashed and solid lines
represent the results with the M3Y-P0, M3Y-P1 and M3Y-P2
interactions, respectively,
while the thin dashed and solid lines those with the SLy5 and D1S
interactions.
\label{fig:NME_M3Ya}}
\end{figure}

In the spin-saturated symmetric nuclear matter,
we have $\eta_s=\eta_t=\eta_{st}=0$,
which indicates $k_{\mathrm{F}p\uparrow}=k_{\mathrm{F}p\downarrow}
=k_{\mathrm{F}n\uparrow}=k_{\mathrm{F}n\downarrow}$
and $\rho_{p\uparrow}=\rho_{p\downarrow}=\rho_{n\uparrow}
=\rho_{n\downarrow}=\rho/4$.
In this case we denote the Fermi momentum simply by $k_\mathrm{F}$.
The lowest energy for a given $\rho$ normally occurs along this line.
The saturation point is obtained
by minimizing the energy per nucleon $\mathcal{E}=E/A$,
\begin{equation}
 \left.\frac{\partial\mathcal{E}}{\partial\rho}\right\vert_\mathrm{sat.}
 =0\,,
\end{equation}
which yields the saturation density $\rho_0$
(equivalently, $k_{\mathrm{F}0}$) and energy $\mathcal{E}_0$.
Figure~\ref{fig:NME_M3Ya} illustrates
$\mathcal{E}$ as a function of $\rho$
for the symmetric nuclear matter with the M3Y-type
as well as with the SLy5 and D1S effective interactions.
We set $M=(M_p+M_n)/2$,
where $M_p$ ($M_n$) is the measured mass of a proton (a neutron).
The parameters for $v_{12}^{(\mathrm{C})}$ and $v_{12}^{(\mathrm{DD})}$
of the M3Y-type interactions
are listed in Table~\ref{tab:param_M3Y}.
As mentioned above, the M3Y-P0 interaction gives no saturation point.
We do have saturation points in M3Y-P1 and M3Y-P2
owing to $v_{12}^{(\mathrm{DD})}$.
Differences among the saturating forces,
\textit{i.e.} SLy5, D1S, M3Y-P1 and M3Y-P2,
are small at $\rho\lesssim \rho_0$.
At relatively high density ($\rho\gtrsim 0.3\,\mathrm{fm}^{-3}$),
the M3Y-P1 and the M3Y-P2 interactions
have lower $\mathcal{E}$ than SLy5 and higher than D1S.
The values of $k_{\mathrm{F}0}$ and $\mathcal{E}_0$ are
tabulated in Table~\ref{tab:NMsat}.
The M3Y-P1 set has been determined so as to give
$k_{\mathrm{F}0}\simeq 1.36\,\mathrm{fm}$
and $\mathcal{E}_0\simeq 16\,\mathrm{MeV}$.

\begin{figure}
\includegraphics[height=16cm]{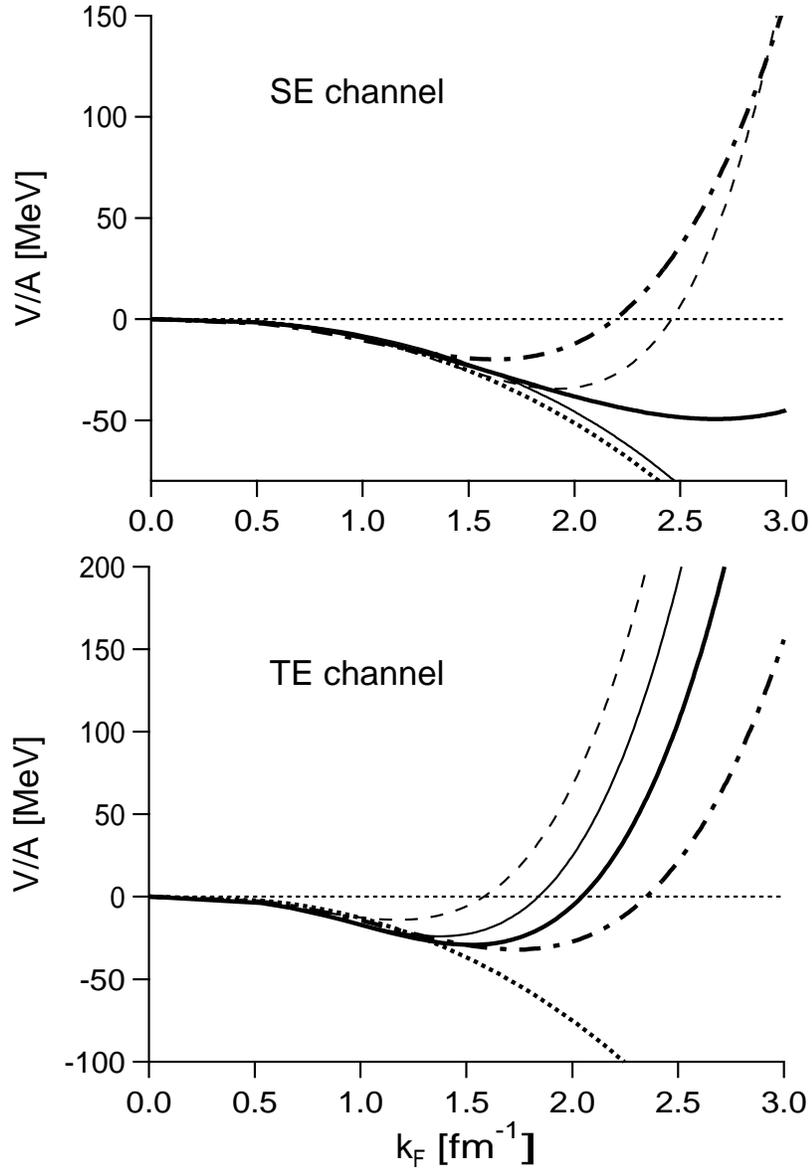}
\vspace{2mm}
\caption{Contribution of the SE and TE channels to $\mathcal{E}$.
See Fig.~\protect\ref{fig:NME_M3Ya} for conventions.
\label{fig:NMV_even}}
\end{figure}

\begin{figure}
\includegraphics[height=16cm]{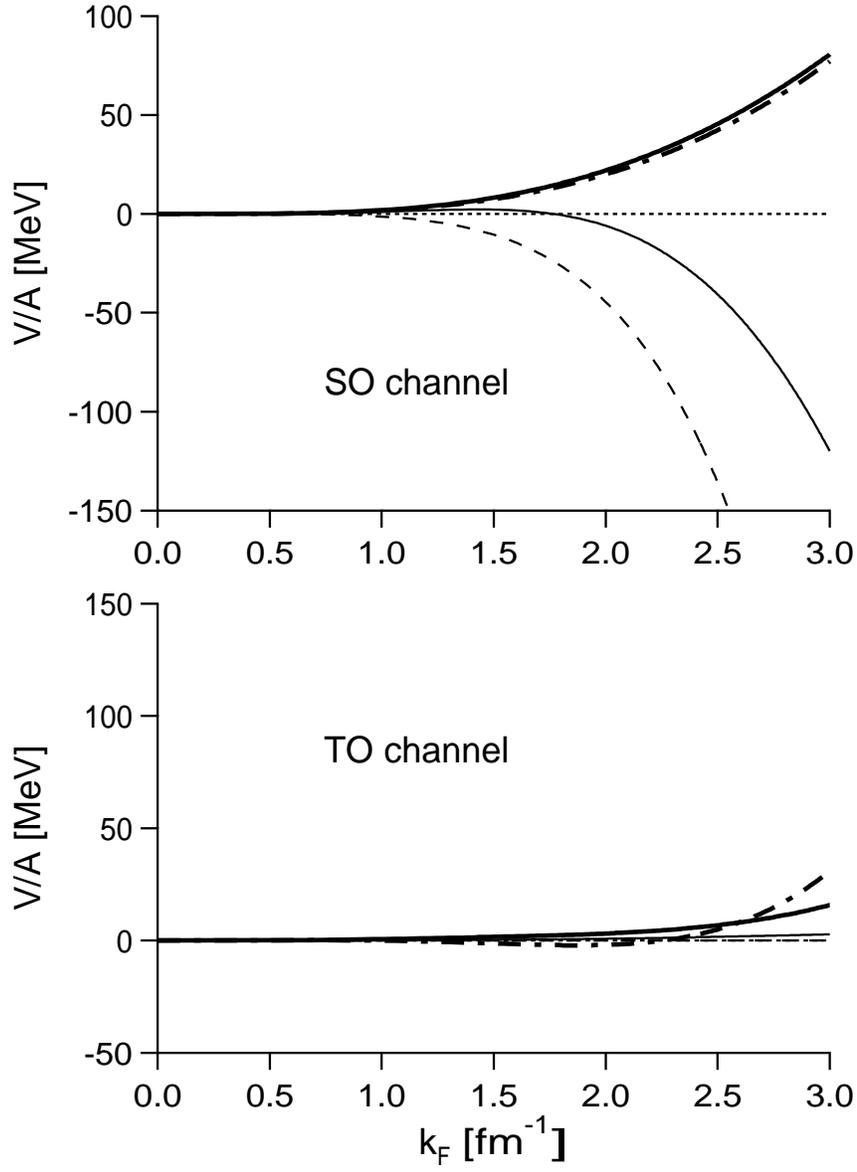}
\vspace{2mm}
\caption{Contribution of the SO and TO channels to $\mathcal{E}$.
In both channels, the results of M3Y-P0 is equal to those of M3Y-P1,
which are presented by the dot-dashed line.
See Fig.~\protect\ref{fig:NME_M3Ya} for the other conventions.
\label{fig:NMV_odd}}
\end{figure}

In Figs.~\ref{fig:NMV_even} and \ref{fig:NMV_odd},
contribution to $\mathcal{E}$ from each of the SE, TE, SO and TO channels
in $v_{12}^{(\mathrm{C})}+v_{12}^{(\mathrm{DD})}$
is shown as a function of $k_\mathrm{F}$.
Sum of all these channels and the kinetic energy
$\langle K\rangle/A = (3/5)(k_\mathrm{F}^2/2M)$
is equal to $\mathcal{E}$ in Fig.~\ref{fig:NME_M3Ya}.
As is viewed in Fig.~\ref{fig:NMV_even},
the TE channel takes a minimum at $k_\mathrm{F}=1.3-1.5\,\mathrm{fm}$
except for M3Y-P0 and M3Y-P1,
primarily responsible for the saturation
at $k_{\mathrm{F}0}\approx 1.3\,\mathrm{fm}$.
In the D1S interaction, the energy out of the SE channel
monotonically goes down.
This is not compatible with the presence
of the strong short-range repulsion in the $NN$ force,
and causes an unphysical property in the neutron matter,
as will be shown in Section~\ref{sec:NMprop2}.
Both the SO and TO channels do not contribute to $\mathcal{E}$
significantly for $\rho\lesssim\rho_0$
(\textit{i.e.} $k_\mathrm{F}\lesssim k_{\mathrm{F}0}$).
While the SO channel becomes attractive
and the TO channel stays small
in the SLy5 and the D1S interactions,
both channels are repulsive in the M3Y-type interactions
at $\rho>\rho_0$, including M3Y-P0.
A certain part of this character of the M3Y-type interactions
comes from the OPEP part.

The curvature at the saturation point with respect to $\rho$
is proportional to the incompressibility,
\begin{equation}
 \mathcal{K} = k_\mathrm{F}^2 \left.\frac{\partial^2\mathcal{E}}
  {\partial k_\mathrm{F}^2}\right\vert_\mathrm{sat.}
 = 9\rho^2 \left.\frac{\partial^2\mathcal{E}}{\partial\rho^2}
       \right\vert_\mathrm{sat.}\,.
\end{equation}
The effective mass ($k$-mass) at the saturation point $M^\ast_0$
is defined by
\begin{equation}
 \left.\frac{\partial\epsilon(\mathbf{k}\sigma\tau)}{\partial k}
 \right\vert_\mathrm{sat.} = \frac{k_{\mathrm{F}0}}{M^\ast_0}\,.
\label{eq:M*}\end{equation}
The volume asymmetry energy corresponds to the curvature of $\mathcal{E}$
with respect to $\eta_t$,
\begin{equation}
 a_t = \left. \frac{1}{2} \frac{\partial^2\mathcal{E}}{\partial\eta_t^2}
	\right\vert_\mathrm{sat.}\,.
\end{equation}
Analogously, the following coefficients are defined
from the curvatures of $\mathcal{E}$
with respect to $\eta_s$ and $\eta_{st}$,
\begin{equation}
 a_s = \left. \frac{1}{2} \frac{\partial^2\mathcal{E}}{\partial\eta_s^2}
	\right\vert_\mathrm{sat.}\,,\quad\quad
 a_{st} = \left. \frac{1}{2} \frac{\partial^2\mathcal{E}}{\partial\eta_{st}^2}
	\right\vert_\mathrm{sat.}\,.
\end{equation}
These coefficients $a_s$, $a_t$ and $a_{st}$ are relevant
to the spin and isospin responses in finite nuclei.
In Table~\ref{tab:NMsat}
we also compare $\mathcal{K}$, $M_0^\ast$, $a_t$, $a_s$ and $a_{st}$
among the effective interactions.

The incompressibility $\mathcal{K}$ is sensitive
to $\alpha$ in $v_{12}^{(\mathrm{DD})}$.
The experimental value of $\mathcal{K}$ has been extracted
from the excitation energies of the giant monopole resonances.
Despite a certain model-dependence,
most non-relativistic models are consistent with the experiments
if $\mathcal{K}\approx 210\,\mathrm{MeV}$.
For finite-range interactions,
\textit{i.e.} the Gogny and the M3Y-type interactions,
$\alpha\approx 1/3$ seems to give reasonable values of $\mathcal{K}$,
while in the Skyrme interactions $\alpha\approx 1/6$ looks favorable,
because of the momentum-dependent terms in $v_{12}^{(\mathrm{C})}$.
The $k$-mass is empirically known
to be $M_0^\ast\approx (0.6-0.7)M$~\cite{ref:Mahaux}.
The M3Y-type interactions tend to yield
slightly smaller $M_0^\ast$
than the SLy5 and the D1S interactions.
The volume asymmetry energy $a_t$ is important
in reproducing global trend of the binding energies
for the $Z\ne N$ nuclei.
From empirical viewpoints $a_t\approx 30\,\mathrm{MeV}$
seems appropriate,
as is fulfilled in the M3Y-type interactions under consideration.

The $a_s$ and $a_{st}$ coefficients are relevant
to the spin degrees of freedom.
The kinetic energy has a certain contribution
to $a_s$ and $a_{st}$, as well as to $a_t$,
which amounts to about $12\,\mathrm{MeV}$
at $\rho\approx\rho_0$ equally for $a_t$, $a_s$ and $a_{st}$.
The interaction $v_{12}^{(\mathrm{C})}+v_{12}^{(\mathrm{DD})}$
gives rise to the rest of these coefficients.
Both the M3Y-type interactions have similar tendency
with respect to these coefficients.
It is remarkable that $a_{st}$ is substantially larger
in the M3Y-type interactions than $a_s$.
As is suggested by close $a_s$ and $a_{st}$ values
between M3Y-P1 and M3Y-P2,
the original M3Y interaction already carries this feature.
In particular, the OPEP part included in the M3Y-type interactions
plays a significant role,
increasing $a_{st}$ by about $11\,\mathrm{MeV}$.
On the other hand, $a_s$ and $a_{st}$ are comparable
in the Gogny D1S interaction,
and we have even $a_s>a_{st}$ in the Skyrme SLy5 interaction.
In the SLy5 case, $a_{st}$ is close to the value
due only to the kinetic energy.

Global characters of the spin and isospin responses
are customarily discussed in terms of the Landau parameters.
Formulae on the Landau parameters at the zero temperature
are given in Appendix~\ref{app:Landau}.
We compute the parameters of Eq.~(\ref{eq:Landau}).
The results are shown in Table~\ref{tab:Landau}.
It is remarked that the M3Y-P1 and M3Y-P2 interactions
give similar results.
The $g_\ell$ and the $g'_\ell$ parameters are closely related
to the $a_s$ and the $a_{st}$ coefficients, respectively.
It has been known that $g_0$ is small,
while $g'_0$ should be relatively large~\cite{ref:g'0}.
Although it is not easy to extract
precise values of the Landau parameters from experimental data
because they could depend on the interaction forms,
qualitative trend will not depend on effective interactions.
The M3Y-type interactions seem to have reasonable characters
on the spin and isospin responses,
while SLy5 and D1S do not,
although the spin and isospin natures of the Skyrme interactions
seem to be improved if the LS currents are ignored~\cite{ref:Lan-Sky}.
It is likely that the difference in these coefficients
may significantly influence predictions of the spin and isospin responses
of finite nuclei.

\section{Properties of asymmetric nuclear matter
and neutron matter\label{sec:NMprop2}}

\begin{figure}
\includegraphics[height=16cm]{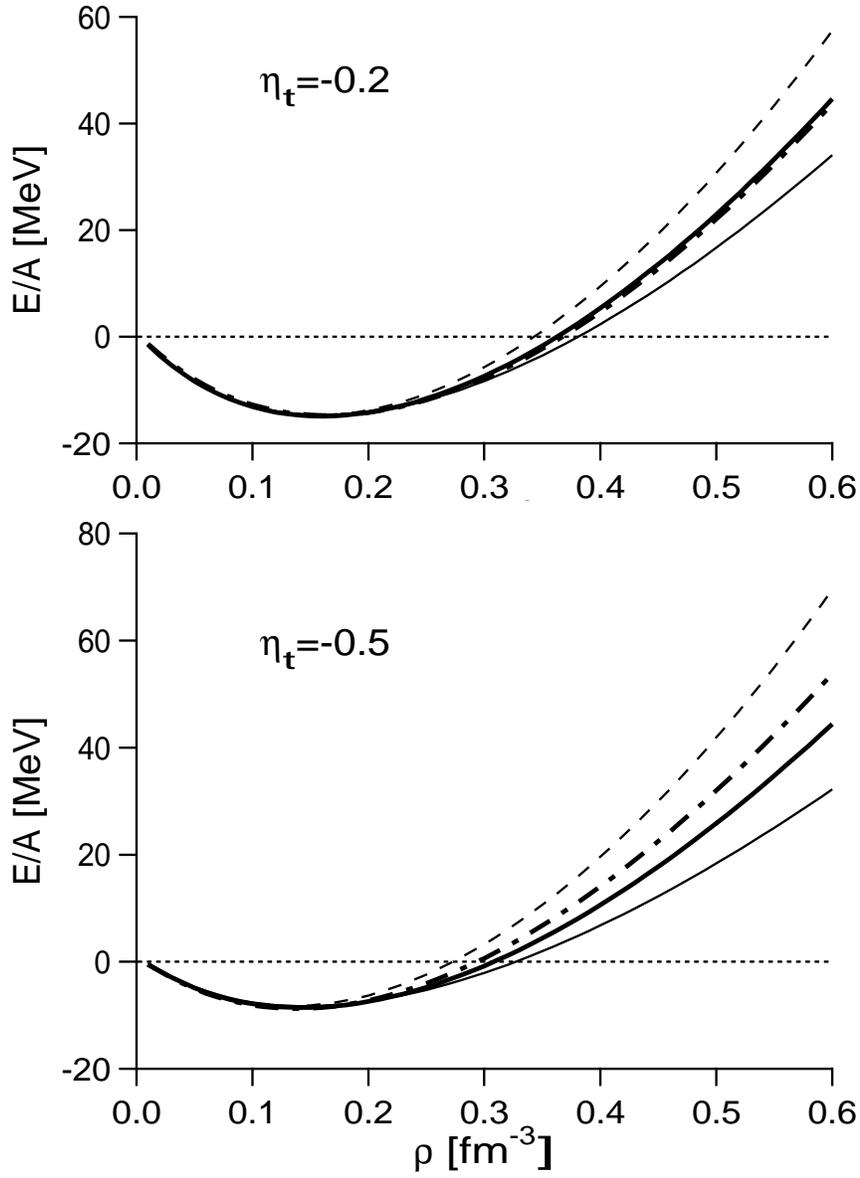}
\vspace{2mm}
\caption{Energies per nucleon $\mathcal{E}=E/A$
in the asymmetric nuclear matters with $\eta_t=-0.2$ and $-0.5$
for several effective interactions.
See Fig.~\protect\ref{fig:NME_M3Ya} for conventions.
\label{fig:NME_M3Yb}}
\end{figure}

\begin{figure}
\includegraphics[height=9cm]{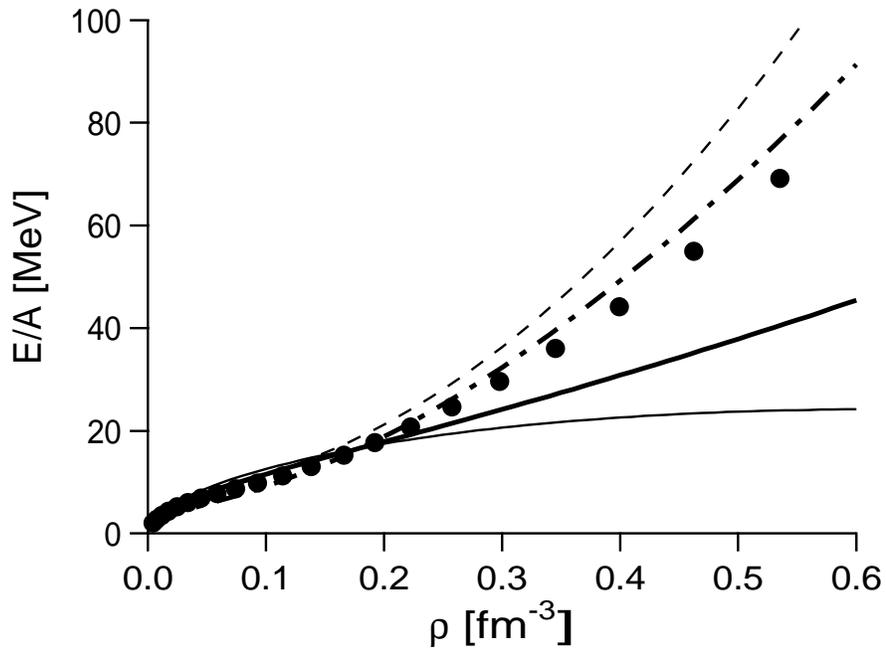}
\vspace{2mm}
\caption{Energies per nucleon $\mathcal{E}=E/A$
in the neutron matter for several effective interactions.
The circles are the results of Ref.~\protect\cite{ref:FP81}.
See Fig.~\protect\ref{fig:NME_M3Ya} for the other conventions.
\label{fig:NME_M3Yc}}
\end{figure}

We turn to the asymmetric nuclear matter.
In Fig.~\ref{fig:NME_M3Yb}, energies per nucleon $\mathcal{E}$
are depicted as a functions of $\rho$
for the spin-saturated (\textit{i.e.} $\eta_s=\eta_{st}=0$)
nuclear matter with $\eta_t=-0.2$ and $-0.5$.
The results from the M3Y-type interactions
are compared with those of the Skyrme and the Gogny interactions.
Energies of the spin-saturated neutron matter
(\textit{i.e.} $\eta_t=-1$) are presented in Fig.~\ref{fig:NME_M3Yc}.
Results from a microscopic calculation in Ref.~\cite{ref:FP81}
are also shown as a reference.
Although the dependence on the interactions is not strong
at low densities even for the neutron matter,
it becomes stronger at $\rho>0.2\,\mathrm{fm}$
as $|\eta_t|$ increases.
In the D1S result for the neutron matter,
$\mathcal{E}$ has a maximum
at $\rho\approx 0.6\,\mathrm{fm}$
and goes to $-\infty$ as $\rho\rightarrow\infty$.
This unphysical behavior arises from $x^{(\mathrm{DD})}=1$
in the D1S set,
which implies no density-dependence in the SE channel
(see Eq.~(\ref{eq:DDchannel})).
This could also give rise to a problem
in practical calculations for finite nuclei.
With the SLy5 interaction $\mathcal{E}$ goes up rapidly
at any $\eta_t$,
because of the momentum-dependence of the interaction.
In contrast to them, the M3Y-type interactions give
moderate $\mathcal{E}$ for the neutron matter.
The microscopic energy of Ref.~\cite{ref:FP81}
lies between those of M3Y-P1 and M3Y-P2.
It will be possible, if necessary,
to adjust the parameters of the M3Y-type interactions
to the microscopic results.

\section{Properties of doubly magic nuclei\label{sec:DMprop}}

We next discuss properties of doubly magic nuclei
in the HF approximation.
In the calculations for finite nuclei,
we use the algorithm presented in Ref.~\cite{ref:NS02},
where the following s.p. bases are employed,
\begin{eqnarray} \varphi_{\alpha\ell jm}(\mathbf{r})
&=& R_{\alpha\ell j}(r)[Y^{(\ell)}(\hat{\mathbf{r}})\chi_\sigma]^{(j)}_m\,;
\nonumber\\
R_{\alpha\ell j}(r) &=& \mathcal{N}_{\alpha\ell j}\,
r^{\ell+2p_\alpha}\exp[-(r/\nu_\alpha)^2]\,.
\label{eq:basis} \end{eqnarray}
Here $Y^{(\ell)}(\hat{\mathbf{r}})$ expresses the spherical harmonics.
We drop the isospin index without confusion.
The index $\alpha$ indicates $p_\alpha$ (a non-negative integer)
and $\nu_\alpha$, simultaneously.
By choosing $p_\alpha$ and $\nu_\alpha$ appropriately,
these bases span the space equivalent to
that of the harmonic-oscillator (HO) bases,
as well as they can form the Kamimura-Gauss (KG) basis-set~\cite{ref:KKF}.
Without parameters specific to mass number or nuclide
such as $\hbar\omega$,
a single set of the KG bases is applicable to wide range of nuclides.
In the following calculations
we apply the hybrid basis-set~\cite{ref:NS02} for the nuclei with $A<50$,
in which an HO basis is added to the KG basis-set,
while the HO basis-set with $N_\mathrm{osc}\leq 15$
and $\hbar\omega=41.2A^{-1/3}\,\mathrm{MeV}$
for heavier nuclei.

In finite nuclei the non-central forces are important as well.
In the M3Y interaction, the LS force $v_{12}^{(\mathrm{LS})}$
and the tensor force $v_{12}^{(\mathrm{TN})}$
are taken by setting $f_n^{(\mathrm{LS})}(r_{12})
=e^{-\mu_n r_{12}}/\mu_n r_{12}$ and $f_n^{(\mathrm{TN})}(r_{12})
=e^{-\mu_n r_{12}}/\mu_n r_{12}$ in Eq.~(\ref{eq:effint}).
We here fix the range parameters as in $v_{12}^{(\mathrm{C})}$;
$\mu_1^{-1}=0.25\,\textrm{fm}$, $\mu_2^{-1}=0.4\,\textrm{fm}$
for $v_{12}^{(\mathrm{LS})}$,
and $\mu_1^{-1}=0.4\,\textrm{fm}$, $\mu_2^{-1}=0.7\,\textrm{fm}$
for $v_{12}^{(\mathrm{TN})}$.
The coupling constants in the M3Y-P2 set
are tabulated in Table~\ref{tab:param_M3Yb},
together with those in the original M3Y-P0 set.
In M3Y-P2,
the enhancement factor for $v_{12}^{(\mathrm{LS})}$ is taken to be 1.8
and the reduction factor for $v_{12}^{(\mathrm{TN})}$ to be 0.12.
The binding energies and the rms matter radii
obtained from the HF calculations with M3Y-P2
are shown in Table~\ref{tab:DMprop},
in comparison with those of the SLy5 and the D1S interactions,
as well as with the experimental data.
The one-body terms of the center-of-mass (c.m.) energy are removed
before iteration.
The contribution of the two-body terms is subtracted
from the convergent HF wave-functions,
in the D1S and the M3Y-P2 results.
There are also spurious c.m. effects
in the matter radii,
\begin{eqnarray}
 \langle r^2\rangle
 &=& \frac{1}{A}\sum_i\langle(\mathbf{r}_i-\mathbf{R})^2\rangle
 ~=~ \frac{1}{A}\sum_i\langle r_i^2\rangle - \langle R^2\rangle
 \nonumber\\
 &=& \frac{1}{A}\left[\left(1-\frac{1}{A}\right)
    \sum_i\langle r_i^2\rangle
  - \frac{1}{A}\sum_{i\ne j}\langle\mathbf{r}_i\cdot\mathbf{r}_j\rangle
    \right]\,.
\label{eq:Rcm}\end{eqnarray}
The first term in the right-hand side is expressed by one-body operators
with a correction factor $(1-1/A)$.
We need two-body operators for the second term.
For the D1S and the M3Y-P2 interactions
we fully remove the c.m. contribution according to Eq.~(\ref{eq:Rcm}).
For the SLy5 interaction we use only the one-body terms
with the correction factor,
ignoring the two-body terms in Eq.~(\ref{eq:Rcm}),
as in calculating the energies.

\begin{figure}
\includegraphics[height=12cm]{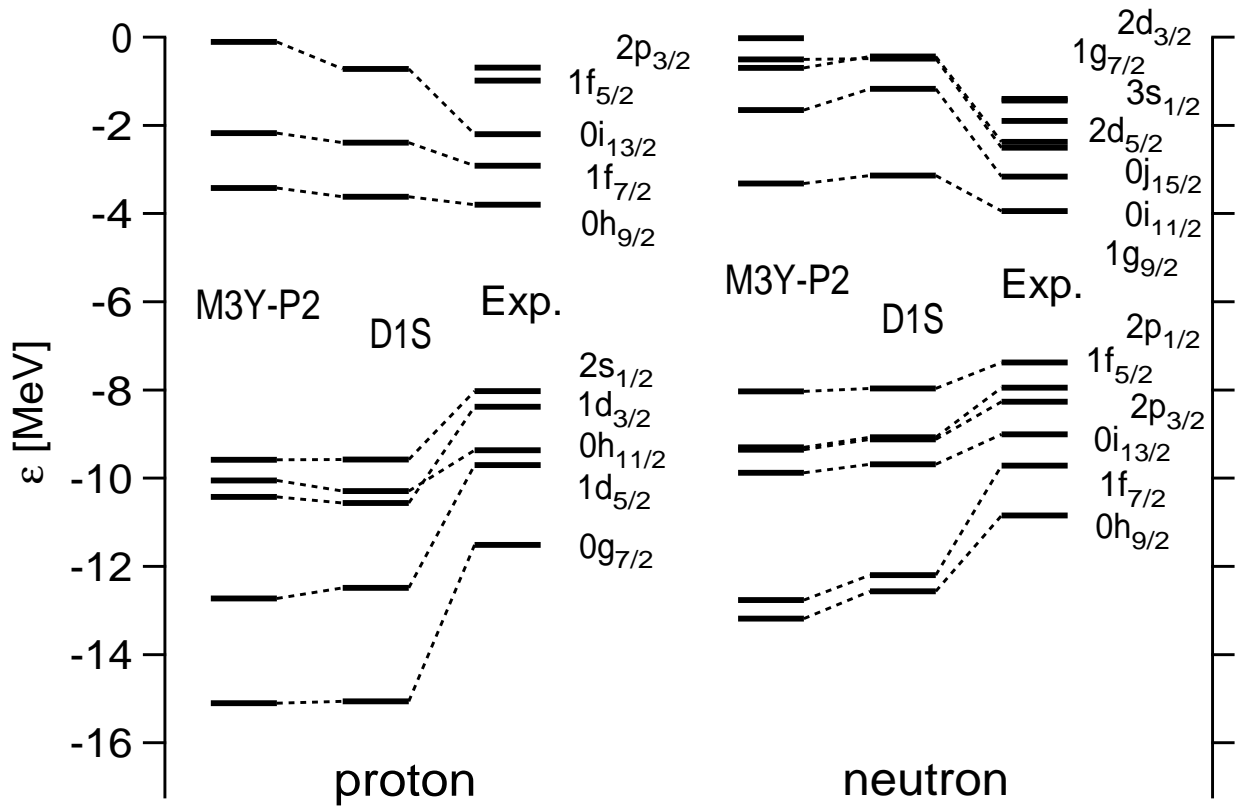}
\vspace{2mm}
\caption{Single-particle energies for $^{208}$Pb.
Experimental values are extracted
from Refs.~\protect\cite{ref:mass,ref:TI}.
\label{fig:Pb_spe}}
\end{figure}

Wave-functions of the doubly magic nuclei are considered
to be well approximated in the spherical HF approaches.
It should still be noted that
correlations due to the residual interaction
could influence their properties.
Therefore we do not pursue fine tuning of the parameters.
As shown in Table~\ref{tab:DMprop}, the M3Y-P2 set is fixed
so as to reproduce the measured binding energies
of the doubly magic nuclei, including $^{90}$Zr,
within about 5\,MeV accuracy.
The binding energies of these nuclei
obtained from the SLy5 and the D1S interactions
are in agreement with the experimental data within 3\,MeV,
slightly better than M3Y-P2.
We do not have to take this difference seriously,
before evaluating influence of the residual interactions.
As well as the binding energies,
the rms matter radii of these nuclei are reproduced
by the M3Y-P2 set similarly well
to the other available interactions.
In Table~\ref{tab:LS} we present the neutron s.p. energies
$\epsilon_n(0p_{3/2})$ and $\epsilon_n(0p_{1/2})$ around $^{16}$O.
The enhancement factor for $v_{12}^{(\mathrm{LS})}$ in the M3Y-P2 set
has been adjusted approximately to the experimental value
of this s.p. energy difference.
The reduction factor for $v_{12}^{(\mathrm{TN})}$ has been determined
so as to reproduce the s.p. energy ordering for $^{208}$Pb.
Without this reduction factor,
the orbits with higher $\ell$ have too high energies.
The resultant s.p. levels in $^{208}$Pb with M3Y-P2
are depicted in Fig.~\ref{fig:Pb_spe}.
The levels obtained from D1S and the experimental s.p. levels
are also shown.
The overall level spacings are related to $M^\ast_0$
shown in Table~\ref{tab:NMsat}.
In the usual HF calculations the level spacings tend to be larger
than the observed ones,
and it is not (and should not be) remedied until the correlations
due to the residual interaction (or the $\omega$-mass)
are taken into account~\cite{ref:Mahaux}.
This is also true in the present case.
We find that M3Y-P2 yields as plausible s.p. levels as D1S does.
We thus confirm that the M3Y-P2 interaction well describes
global nature of stable nuclei.

\section{Single particle levels in $N=16$ isotones
\label{sec:N16}}

In the preceding section
we have shown that the M3Y-P2 interaction reproduces
the properties of the doubly magic nuclei
to a similar accuracy to the SLy5 and the D1S interactions.
At a glance, the spin-isospin characters in the nuclear matter,
which have been discussed in Section~\ref{sec:NMprop}
via $a_{st}$ and $g'_\ell$,
do not seem to influence the nuclear properties
around the ground states.
However, the spin and isospin characters
influence s.p. energies of finite nuclei.
Thereby they may affect even the ground state properties.
In this section we illustrate this point
by the neutron orbits in the $N=16$ isotones,
following the arguments in Ref.~\cite{ref:Vst},
although precise studies in this line are beyond scope of this paper.

\begin{figure}
\includegraphics[height=10cm]{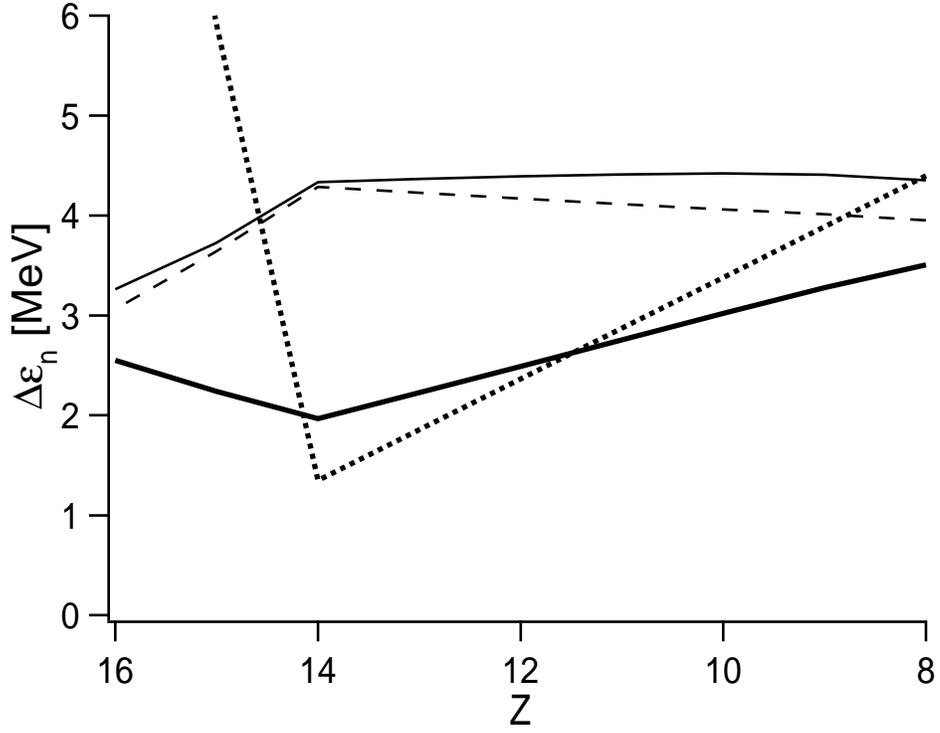}
\vspace{2mm}
\caption{${\mathit\Delta}\epsilon_n$ for the $N=16$ isotones.
The thick solid, dotted, thin solid and dashed lines
correspond to the results with the M3Y-P2, USD, D1S and SLy5
interactions, respectively.
\label{fig:dspe_N16}}
\end{figure}

As was suggested in Ref.~\cite{ref:Vst},
the proton-number ($Z$) dependence of the neutron s.p. energy
$\epsilon_n(0d_{3/2})$ relative to $\epsilon_n(1s_{1/2})$
can sizably be affected by effective interactions.
Figure~\ref{fig:dspe_N16} depicts
${\mathit\Delta}\epsilon_n=\epsilon_n(0d_{3/2})-\epsilon_n(1s_{1/2})$
obtained from the spherical HF calculations in the $N=16$ isotones.
Though it is not obvious
whether the ground states of all of these isotones
are well approximated by the spherical HF wave-functions,
it is meaningful to see the s.p. energies,
which often give an indication to magic or submagic numbers.
For D1S we reduce the number of bases in Eq.~(\ref{eq:basis})
to avoid instability occurring for some $N=16$ nuclei,
which probably relates to the unphysical behavior
in the neutron matter.
It is found that, if viewed as a function of $Z$,
${\mathit\Delta}\epsilon_n$ strikingly depends on the interactions.
With the M3Y-P2 interaction, ${\mathit\Delta}\epsilon_n$ increases
as $Z$ goes from $Z=14$ to $Z=8$.
We have confirmed~\cite{ref:Nak02}
that even M3Y-P1 (with appropriate $v_{12}^{(\mathrm{LS})}$
and $v_{12}^{(\mathrm{TN})}$) shows similar behavior
and that a significant part of this feature
originates in the OPEP part in $v_{12}^{(\mathrm{C})}$.
It is thus suggested that
this behavior of ${\mathit\Delta}\epsilon_n$ is correlated
to the spin-isospin property in the nuclear matter.

For comparison, we also show the s.p. energies
obtained from the reliable shell model interaction
for the $sd$-shell nuclei,
the so-called USD interaction~\cite{ref:USD}.
For this purpose we define
effective values of the s.p. energies for each nucleus $A$
from the shell model space and interaction,
which correspond to those of the spherical HF calculations,
as
\begin{equation}
 \epsilon_n^\mathrm{USD}(j;A)
  = \epsilon_n^\mathrm{USD}(j;^{\,17\!}\mathrm{O})
  + \sum_{j'} \langle N_{j'}\rangle_A\, \frac{2J+1}{(2j+1)(2j'+1)}\,
  \langle jj';J|v^\mathrm{USD}|jj';J\rangle\,,
\end{equation}
where the sum with respect to $j'$ runs over
the valence orbits.
For $\langle N_j\rangle_A$,
we assume that the nucleons occupy the s.p. orbits
from the bottom, according to $\epsilon(j)$.
From these s.p. energies we obtain
${\mathit\Delta}\epsilon_n^\mathrm{USD}
=\epsilon_n^\mathrm{USD}(0d_{3/2};A)
-\epsilon_n^\mathrm{USD}(1s_{1/2};A)$
for individual nucleus.
This definition is equivalent to the effective s.p. energies
in Ref.~\cite{ref:Vst} for the $Z\leq N(=16)$ nuclei.
The ${\mathit\Delta}\epsilon_n^\mathrm{USD}$ values
are also shown in Fig.~\ref{fig:dspe_N16}.
It is noted that in the shell model approaches
the nucleus-dependence of the s.p. wave-functions
is not fully taken into account.
Effects of rearrangement in the wave-functions of the deeply bound orbits
are renormalized into the interactions among the valence nucleons.
In contrast, in the HF approaches
the s.p. wave-functions are determined self-consistently,
from nucleus to nucleus.
Therefore the shell model s.p. energies
do not agree with their HF counterparts.
However, there should be qualitative correspondence,
which arises from basic characters of the effective interactions.
It is remarked that the M3Y-P2 interaction
has the same trend of ${\mathit\Delta}\epsilon_n$,
in terms of the $Z$-dependence, as the USD interaction.
It has been suggested~\cite{ref:Vst} that the interaction
in the $(\mbox{\mathversion{bold}$\sigma$}\cdot
\mbox{\mathversion{bold}$\sigma$})\,
(\mbox{\mathversion{bold}$\tau$}\cdot
\mbox{\mathversion{bold}$\tau$})$ channel,
which will be linked to $a_{st}$ or to $g'_\ell$,
is significant to the magic numbers in highly neutron-rich nuclei,
and that the $Z$-dependence of the s.p. energies
in this region could be relevant to the new magic number
$N=16$~\cite{ref:N16}.
The present results are fully consistent
with the arguments in Ref.~\cite{ref:Vst},
although we cannot draw conclusions on the magic number problem
without assessing influence of the residual interactions.

\section{Summary and outlook\label{sec:summary}}

We have developed effective interactions
to describe low energy phenomena of nuclei.
Starting from the M3Y interaction,
we introduce a density-dependent contact term
and modify several parameters in a phenomenological manner,
whereas maintaining the OPEP part in the central force.
In order to view basic characters of the interactions,
the Hartree-Fock calculations are implemented
for the infinite nuclear matter
(for which useful formulae are newly derived)
and for several doubly magic nuclei.
We have shown that a parameter-set called M3Y-P2
describes their properties plausibly.
The properties which are well treated by the Skyrme SLy5
and/or the Gogny D1S interactions
are also reproduced by the M3Y-P2 interaction.
However, a remarkable difference is found
in the properties relevant to the spin degrees of freedom
in the nuclear matter.
The M3Y-type interactions seem to give
reasonable spin and isospin properties,
in which the OPEP part contained in $v_{12}^{(\mathrm{C})}$
plays a significant role.
We have also shown that the difference in the spin-isospin property
affects the s.p. energies in finite nuclei to a considerable extent.
It will be interesting to apply extensively the M3Y-type interactions,
particularly to the magic number problems far from the $\beta$-stability.

Although the M3Y-P2 interaction seems to have various desired characters,
there still remains a certain room for further tuning of the parameters.
It should be noted that this parameter-set will not be
a unique choice to reproduce the properties of the nuclear matter
and the doubly magic nuclei.
Effective interaction might not be constrained sufficiently
only from the HF calculations.
The pairing effects in nuclei give valuable information
on the effective interaction,
primarily on the SE channel.
Comparison of the matrix elements
with reliable shell model interactions will also be helpful,
if the core polarization effects are treated appropriately.
These points will be discussed in future publications.

\clearpage
\begin{table}
\begin{center}
\caption{Parameters of central forces (including $v_{12}^{(\mathrm{DD})}$)
 in the original and modified M3Y interactions.
 See text for the $\mu_n$ parameters.
\label{tab:param_M3Y}}
\begin{tabular}{ccr@{.}lr@{.}lr@{.}l}
\hline\hline
parameters && \multicolumn{2}{c}{~~M3Y-P0~~} &
 \multicolumn{2}{c}{~~M3Y-P1~~} & \multicolumn{2}{c}{~~M3Y-P2~~}
 \\ \hline
$t_1^{(\mathrm{SE})}$ &(MeV)& $11466$& & $8599$&$5$ & $8027$& \\
$t_1^{(\mathrm{TE})}$ &(MeV)& $13967$& & $10475$&$25$ & $6080$& \\
$t_1^{(\mathrm{SO})}$ &(MeV)& $-1418$& & $-1418$& & $-11900$& \\
$t_1^{(\mathrm{TO})}$ &(MeV)& $11345$& & $11345$& & $3800$& \\
$t_2^{(\mathrm{SE})}$ &(MeV)& $-3556$& & $-3556$& & $-2880$& \\
$t_2^{(\mathrm{TE})}$ &(MeV)& $-4594$& & $-4594$& & $-4266$& \\
$t_2^{(\mathrm{SO})}$ &(MeV)& $950$& & $950$& & $2730$& \\
$t_2^{(\mathrm{TO})}$ &(MeV)& $-1900$& & $-1900$& & $-780$& \\
$t_3^{(\mathrm{SE})}$ &(MeV)& $-10$&$463$ & $-10$&$463$ & $-10$&$463$ \\
$t_3^{(\mathrm{TE})}$ &(MeV)& $-10$&$463$ & $-10$&$463$ & $-10$&$463$ \\
$t_3^{(\mathrm{SO})}$ &(MeV)& $31$&$389$ & $31$&$389$ & $31$&$389$ \\
$t_3^{(\mathrm{TO})}$ &(MeV)& $3$&$488$ & $3$&$488$ & $3$&$488$ \\
$\alpha$ && \multicolumn{2}{c}{---} &
 \multicolumn{2}{c}{$1/3$}& \multicolumn{2}{c}{$1/3$} \\
$t^{(\mathrm{DD})}$ &(MeV$\cdot$fm)& $0$& & $1212$& & $1320$& \\
$x^{(\mathrm{DD})}$ && \multicolumn{2}{c}{---} &
 \multicolumn{2}{r}{$0.09834$}& \multicolumn{2}{r}{$0.72576$} \\
\hline\hline
\end{tabular}
\end{center}
\end{table}

\begin{table}
\begin{center}
\caption{Nuclear matter properties at the saturation point.
\label{tab:NMsat}}
\begin{tabular}{ccrrrr}
\hline\hline
&&~~M3Y-P1 &~~M3Y-P2 &~~~~SLy5~ &~~~~D1S~~ \\ \hline
$k_{\mathrm{F}0}$ & (fm) & $1.358$~~& $1.340$~~&
 $1.334$~~& $1.342$~~\\
$\mathcal{E}_0$ & (MeV) & $-15.99$~~& $-16.14$~~&
 $-15.98$~~& $-16.01$~~\\
$\mathcal{K}$ & (MeV) & $225.7$~~& $220.4$~~&
 $229.9$~~& $202.9$~~\\
$M^\ast_0/M$ && $0.641$~~& $0.652$~~&
 $0.697$~~& $0.697$~~\\
$a_t$ & (MeV) & $30.35$~~& $30.61$~~&
 $32.03$~~& $31.12$~~\\
$a_s$ & (MeV) & $20.81$~~& $21.19$~~&
 $37.47$~~& $26.18$~~\\
$a_{st}$ & (MeV) & $37.63$~~& $38.19$~~&
 $15.15$~~& $29.13$~~\\
\hline\hline
\end{tabular}
\end{center}
\end{table}

\begin{table}
\begin{center}
\caption{Landau parameters at the saturation point.
\label{tab:Landau}}
\begin{tabular}{cr@{.}lr@{.}lr@{.}lr@{.}l}
\hline\hline
\hspace*{1cm} &\multicolumn{2}{c}{M3Y-P1~~} &
 \multicolumn{2}{c}{M3Y-P2~~} &\multicolumn{2}{c}{~~SLy5~~~} &
 \multicolumn{2}{c}{~~D1S~~~~} \\ \hline
$f_0$ & $-0$&$370$ & $-0$&$357$ & $-0$&$276$ & $-0$&$369$ \\
$f_1$ & $-1$&$078$ & $-1$&$044$ & $-0$&$909$ & $-0$&$909$ \\
$f_2$ & $-0$&$381$ & $-0$&$436$ & $0$&$0$ & $-0$&$558$ \\
$f_3$ & $-0$&$191$ & $-0$&$210$ & $0$&$0$ & $-0$&$157$ \\ \hline
$f'_0$ & $0$&$525$ & $0$&$607$ & $0$&$815$ & $0$&$743$ \\
$f'_1$ & $0$&$537$ & $0$&$635$ & $-0$&$387$ & $0$&$470$ \\
$f'_2$ & $0$&$250$ & $0$&$245$ & $0$&$0$ & $0$&$342$ \\
$f'_3$ & $0$&$101$ & $0$&$096$ & $0$&$0$ & $0$&$100$ \\ \hline
$g_0$ & $0$&$046$ & $0$&$113$ & $1$&$123$ & $0$&$466$ \\
$g_1$ & $0$&$372$ & $0$&$273$ & $0$&$253$ & $-0$&$184$ \\
$g_2$ & $0$&$199$ & $0$&$162$ & $0$&$0$ & $0$&$245$ \\
$g_3$ & $0$&$088$ & $0$&$078$ & $0$&$0$ & $0$&$091$ \\ \hline
$g'_0$ & $0$&$891$ & $1$&$006$ & $-0$&$141$ & $0$&$631$ \\
$g'_1$ & $0$&$230$ & $0$&$202$ & $1$&$043$ & $0$&$610$ \\
$g'_2$ & $0$&$073$ & $0$&$040$ & $0$&$0$ & $-0$&$038$ \\
$g'_3$ & $0$&$008$ & $-0$&$002$ & $0$&$0$ & $-0$&$036$ \\
\hline\hline
\end{tabular}
\end{center}
\end{table}

\begin{table}
\begin{center}
\caption{Parameters of non-central forces
 in the original and modified M3Y interactions.
 See text for the $\mu_n$ parameters.
\label{tab:param_M3Yb}}
\begin{tabular}{ccr@{.}lr@{.}l}
\hline\hline
parameters && \multicolumn{2}{c}{M3Y-P0~~\quad} &
 \multicolumn{2}{c}{M3Y-P2~~\quad} \\ \hline
$t_1^{(\mathrm{LSE})}$ &(MeV)&\quad $-5101$& &\quad $-9181$&$8$ \\
$t_1^{(\mathrm{LSO})}$ &(MeV)& $-1897$& & $-3414$&$6$ \\
$t_2^{(\mathrm{LSE})}$ &(MeV)& $-337$& & $-606$&$6$ \\
$t_2^{(\mathrm{LSO})}$ &(MeV)& $-632$& & $-1137$&$6$ \\
$t_1^{(\mathrm{TNE})}$ &(MeV)& $-1096$& & $-131$&$52$ \\
$t_1^{(\mathrm{TNO})}$ &(MeV)& $244$& & $29$&$28$ \\
$t_2^{(\mathrm{TNE})}$ &(MeV)& $-30$&$9$ & $-3$&$708$ \\
$t_2^{(\mathrm{TNO})}$ &(MeV)& $15$&$6$ & $1$&$872$ \\
\hline\hline
\end{tabular}
\end{center}
\end{table}

\begin{table}
\begin{center}
\caption{Binding energies and rms matter radii
 of several doubly magic nuclei.
 Experimental data are taken
 from Refs.~\protect\cite{ref:mass,ref:O16-rad,ref:rad}.
\label{tab:DMprop}}
\begin{tabular}{cccrrrr}
\hline\hline
&&&~~~Exp.~~& M3Y-P2~&~~~SLy5~~&~~~D1S~~\\ \hline
$^{16}$O & $-E$ &(MeV)&
 $127.6$ & $127.1$ & $128.6$ & $129.5$ \\
& $\sqrt{\langle r^2\rangle}$ &(fm)&
 $2.61$ & $2.60$ & $2.59$ & $2.59$ \\
$^{40}$Ca & $-E$ &(MeV)&
 $342.1$ & $338.7$ & $344.3$ & $344.5$ \\
& $\sqrt{\langle r^2\rangle}$ &(fm)&
 $3.47$ & $3.37$ & $3.29$ & $3.36$ \\
$^{48}$Ca & $-E$ &(MeV)&
 $416.0$ & $411.8$ & $416.0$ & $416.8$ \\
& $\sqrt{\langle r^2\rangle}$ &(fm)&
 $3.57$ & $3.52$ & $3.44$ & $3.50$ \\
$^{90}$Zr & $-E$ &(MeV)&
 $783.9$ & $778.7$ & $782.4$ & $784.5$ \\
& $\sqrt{\langle r^2\rangle}$ &(fm)&
 $4.32$ & $4.25$ & $4.22$ & $4.23$ \\
$^{132}$Sn & $-E$ &(MeV)&
 $1102.9$ & $1098.1$ & $1103.5$ & $1102.9$ \\
& $\sqrt{\langle r^2\rangle}$ &(fm)&
 --- & $4.79$ & $4.77$ & $4.76$ \\
$^{208}$Pb & $-E$ &(MeV)&
 $1636.4$ & $1635.8$ & $1635.2$ & $1638.1$ \\
& $\sqrt{\langle r^2\rangle}$ &(fm)&
 $5.49$ & $5.53$ & $5.52$ & $5.51$ \\
\hline\hline
\end{tabular}
\end{center}
\end{table}

\begin{table}
\begin{center}
\caption{LS splitting around $^{16}$O.
 Experimental data are extracted from Refs.~\protect\cite{ref:mass,ref:TI}.
\label{tab:LS}}
\begin{tabular}{ccrrrr}
\hline\hline
&&~~~Exp.~~& M3Y-P2~&~~~SLy5~~&~~~D1S~~\\ \hline
$\epsilon_n(0p_{3/2})$ &(MeV)&
 $-21.8$ & $-22.6$ & $-20.6$ & $-22.3$ \\
$\epsilon_n(0p_{1/2})$ &(MeV)&
 $-15.7$ & $-16.2$ & $-14.4$ & $-15.9$ \\
\hline\hline
\end{tabular}
\end{center}
\end{table}

\clearpage

\clearpage

\appendix
\section{Analytic formulae for nuclear matter energy\label{app:NME}}

In this Appendix we derive formulae
concerning the interaction part of Eq.~(\ref{eq:NME}).
The form of Eq.~(\ref{eq:effint}) is assumed for $v_{12}$.

Each term of $v_{12}^{(\mathrm{C})}$ is expressed
as $f_n^{(\mathrm{C})}(r_{12})\mathcal{O}_\sigma \mathcal{O}_\tau$.
Its non-antisymmetrized matrix element
in the plane wave states of Eq.~(\ref{eq:NM-spwf})
is evaluated as
\begin{eqnarray}
&& \langle \mathbf{k}_1'\sigma_1'\tau_1',\mathbf{k}_2'\sigma_2'\tau_2'
|f_n^{(\mathrm{C})}(r_{12}) \mathcal{O}_\sigma \mathcal{O}_\tau|
\mathbf{k}_1\sigma_1\tau_1,\mathbf{k}_2\sigma_2\tau_2 \rangle_\mathrm{n.a.}
\nonumber\\
&&= \frac{1}{\Omega^2} \int d^3r_1 d^3r_2\,
e^{i(\mathbf{k}_1-\mathbf{k}_1')\cdot\mathbf{r}_1
+i(\mathbf{k}_2-\mathbf{k}_2')\cdot\mathbf{r}_2}\,
f_n^{(\mathrm{C})}(r_{12})\,
\langle \sigma_1'\sigma_2'|\mathcal{O}_\sigma
|\sigma_1\sigma_2\rangle\,
\langle \tau_1'\tau_2'|\mathcal{O}_\tau|\tau_1\tau_2\rangle \nonumber\\
&&= \frac{1}{\Omega^2} \int d^3R d^3r_{12}\,
e^{i(\mathbf{K}-\mathbf{K}')\cdot\mathbf{R}
+i(\mathbf{k}_{12}-\mathbf{k}_{12}')\cdot\mathbf{r}_{12}}\,
f_n^{(\mathrm{C})}(r_{12})\,
\langle \sigma_1'\sigma_2'|\mathcal{O}_\sigma
|\sigma_1\sigma_2\rangle\,
\langle \tau_1'\tau_2'|\mathcal{O}_\tau|\tau_1\tau_2\rangle \nonumber\\
&&= \frac{1}{\Omega}\,\delta_{\mathbf{K},\mathbf{K}'}\,
\tilde f_n^{(\mathrm{C})}(|\mathbf{k}_{12}-\mathbf{k}_{12}'|)\,
\langle \sigma_1'\sigma_2'|\mathcal{O}_\sigma
|\sigma_1\sigma_2\rangle\,
\langle \tau_1'\tau_2'|\mathcal{O}_\tau|\tau_1\tau_2\rangle\,,
\label{eq:NME-tbme}\end{eqnarray}
where $\mathbf{R}=(\mathbf{r}_1+\mathbf{r}_2)/2$,
$\mathbf{r}_{12}=\mathbf{r}_1-\mathbf{r}_2$,
$\mathbf{K}=\mathbf{k}_1+\mathbf{k}_2$,
$\mathbf{K}'=\mathbf{k}_1'+\mathbf{k}_2'$,
$\mathbf{k}_{12}=(\mathbf{k}_1-\mathbf{k}_2)/2$,
$\mathbf{k}_{12}'=(\mathbf{k}_1'-\mathbf{k}_2')/2$
and $\tilde f(q)$ is the Fourier transform of $f(r)$,
\begin{equation} \tilde f (q)
= \int d^3r\,f(r) e^{-i\mathbf{q}\cdot\mathbf{r}}\,.
\label{eq:Four1}\end{equation}
The density-dependent interaction $v_{12}^{(\mathrm{DD})}$
is also handled in a similar manner,
since the density behaves like a constant in the nuclear matter.
For the Hartree term we have
$(\mathbf{k}_1\sigma_1\tau_1)=(\mathbf{k}_1'\sigma_1'\tau_1')$
and $(\mathbf{k}_2\sigma_2\tau_2)=(\mathbf{k}_2'\sigma_2'\tau_2')$,
while $(\mathbf{k}_1\sigma_1\tau_1)=(\mathbf{k}_2'\sigma_2'\tau_2')$
and $(\mathbf{k}_2\sigma_2\tau_2)=(\mathbf{k}_1'\sigma_1'\tau_1')$
for the Fock term.
Therefore both terms satisfy $\mathbf{K}=\mathbf{K}'$.
For the relative momentum
the Hartree term (the Fock term) yields
$\mathbf{k}_{12}-\mathbf{k}_{12}'=0$
($\mathbf{k}_{12}-\mathbf{k}_{12}'=2\mathbf{k}_{12}$).
Contribution of the two-body interaction to the nuclear matter energy
is obtained by integrating $\tilde f$ in Eq.~(\ref{eq:NME-tbme})
up to the Fermi momenta.

We here consider general cases
where the Fermi momentum may depend on spin and isospin.
In order to take into account the spin-isospin dependence,
we integrate $\tilde f$ in the range $k_1\leq k_{\mathrm{F}1}$
and $k_2\leq k_{\mathrm{F}2}$.
The integration is immediately carried out for the Hartree term,
as far as $f(r_{12})$ is momentum-independent,
since the integrand depends neither on $\mathbf{k}_1$
nor on $\mathbf{k}_2$,
\begin{equation}
\mathcal{W}^\mathrm{H}(k_{\mathrm{F}1},k_{\mathrm{F}2})
= \int_{k_1\leq k_{\mathrm{F}1}} d^3k_1
\int_{k_2\leq k_{\mathrm{F}2}} d^3k_2\,
\tilde f(0)
= \frac{16\pi^2}{9} k_{\mathrm{F}1}^3 k_{\mathrm{F}2}^3\,
\tilde f(0)\,.
\end{equation}
For the Fock term contribution,
the integral with respect to $\mathbf{k}_1$ and $\mathbf{k}_2$
is converted to the one
with respect to $\mathbf{K}$ and $\mathbf{k}_{12}$.
We here assume $k_{\mathrm{F}1}\leq k_{\mathrm{F}2}$
without loss of generality, owing to the symmetry
$\mathcal{W}(k_{\mathrm{F}1},k_{\mathrm{F}2})
= \mathcal{W}(k_{\mathrm{F}2},k_{\mathrm{F}1})$.
Handling the range of integral carefully,
we obtain the following expression,
\begin{eqnarray}
\mathcal{W}^\mathrm{F}(k_{\mathrm{F}1},k_{\mathrm{F}2})
&=& \int_{k_1\leq k_{\mathrm{F}1}} d^3k_1
\int_{k_2\leq k_{\mathrm{F}2}} d^3k_2\,
\tilde f(2k_{12}) \nonumber\\
&=& 8\pi^2 \left[ \int_0^{(k_{\mathrm{F}2}-k_{\mathrm{F}1})/2}
dk_{12}\,\frac{16}{3}k_{\mathrm{F}1}^3 k_{12}^2\,
\tilde f(2k_{12}) \right. \nonumber\\
&&\quad \left. + \int_{(k_{\mathrm{F}2}-k_{\mathrm{F}1})/2}
^{(k_{\mathrm{F}1}+k_{\mathrm{F}2})/2}
dk_{12}\,\left\{-\frac{1}{2}(k_{\mathrm{F}2}^2-k_{\mathrm{F}1}^2)^2 k_{12}
+\frac{8}{3}(k_{\mathrm{F}1}^3+k_{\mathrm{F}2}^3) k_{12}^2
\right.\right.
\nonumber\\
&&\hspace*{4cm} \left.\left.
-4(k_{\mathrm{F}1}^2+k_{\mathrm{F}2}^2) k_{12}^3
+\frac{8}{3}k_{12}^5 \right\}\,
\tilde f(2k_{12})
\right]
\end{eqnarray}
These formulae are general to multi-component uniform Fermi liquids
with equal masses.

In handling the spin-isospin degrees of freedom,
we rewrite the central force in Eq.~(\ref{eq:effint}) as
\begin{equation}
v_{12}^{(\mathrm{C})} = \sum_n (t_n^{(\mathrm{W})}
 + t_n^{(\mathrm{B})} P_\sigma - t_n^{(\mathrm{H})} P_\tau
 - t_n^{(\mathrm{M})} P_\sigma P_\tau)
 f_n^{(\mathrm{C})} (r_{12})\,.
\end{equation}
The relations between the coupling constants are
\begin{eqnarray}
t_n^{(\mathrm{SE})} = t_n^{(\mathrm{W})}-t_n^{(\mathrm{B})}
-t_n^{(\mathrm{H})}+t_n^{(\mathrm{M})}\,,&&
t_n^{(\mathrm{TE})} = t_n^{(\mathrm{W})}+t_n^{(\mathrm{B})}
+t_n^{(\mathrm{H})}+t_n^{(\mathrm{M})}\,,\nonumber\\
t_n^{(\mathrm{SO})} = t_n^{(\mathrm{W})}-t_n^{(\mathrm{B})}
+t_n^{(\mathrm{H})}-t_n^{(\mathrm{M})}\,,&&
t_n^{(\mathrm{TO})} = t_n^{(\mathrm{W})}+t_n^{(\mathrm{B})}
-t_n^{(\mathrm{H})}-t_n^{(\mathrm{M})}\,.
\end{eqnarray}
After summing over the spin-isospin degrees of freedom,
the interaction energy is given by
\begin{eqnarray} \langle V\rangle
&=& \frac{\Omega}{2(2\pi)^6} \sum_n \sum_{\sigma_1\sigma_2\tau_1\tau_2}
\left[\left(t_n^{(\mathrm{W})}
+ t_n^{(\mathrm{B})} \delta_{\sigma_1\sigma_2}
- t_n^{(\mathrm{H})} \delta_{\tau_1\tau_2}
- t_n^{(\mathrm{M})} \delta_{\sigma_1\sigma_2} \delta_{\tau_1\tau_2}
\right) \mathcal{W}_n^\mathrm{H}(k_{\mathrm{F}\tau_1\sigma_1},
k_{\mathrm{F}\tau_2\sigma_2})
\right. \nonumber\\
&&\hspace*{2cm} \left. + \left(t_n^{(\mathrm{M})}
+ t_n^{(\mathrm{H})} \delta_{\sigma_1\sigma_2}
- t_n^{(\mathrm{B})} \delta_{\tau_1\tau_2}
- t_n^{(\mathrm{W})} \delta_{\sigma_1\sigma_2} \delta_{\tau_1\tau_2}
\right) \mathcal{W}_n^\mathrm{F}(k_{\mathrm{F}\tau_1\sigma_1},
k_{\mathrm{F}\tau_2\sigma_2}) \right]\,.
\label{eq:NME-tot}\end{eqnarray}
In Eq.~(\ref{eq:NME-tot}) we regard the sum over $n$
to include $v_{12}^{(\mathrm{DD})}$.
It is noted that $A=\Omega\rho$, which is used
to obtain the energy per nucleon $\mathcal{E}$.

We next calculate the $\mathcal{W}$ functions
for typical interaction forms.
\begin{enumerate}
 \item \textit{$\delta$ interaction}\\
If $f(r_{12})=\delta(\mathbf{r}_{12})$,
$\tilde f(q)=1$ and therefore we have
\begin{equation}
\mathcal{W}^\mathrm{H}(k_1,k_2) = \mathcal{W}^\mathrm{F}(k_1,k_2)
= \frac{16\pi^2}{9} k_1^3 k_2^3\,.
\end{equation}
 \item \textit{$\rho$-dependent $\delta$ interaction}\\
Since the density is a constant in the uniform nuclear matter,
the $\mathcal{W}$ functions for $f(r_{12})
= \rho^\alpha \delta({\mathbf{r}_{12}})$
are similar to the above case,
\begin{equation}
\mathcal{W}^\mathrm{H}(k_1,k_2) = \mathcal{W}^\mathrm{F}(k_1,k_2)
= \frac{16\pi^2}{9} \rho^\alpha k_1^3 k_2^3\,.
\end{equation}
Note that $\rho$ is a function of the Fermi momenta,
when we take derivatives of the $\mathcal{W}$ functions.
 \item \textit{Gauss interaction}\\
For $f(r_{12})=e^{-(\mu r_{12})^2}$,
we have $\tilde f(q)=({\sqrt\pi}/\mu)^3\, e^{-(q/2\mu)^2}$,
deriving
\begin{equation}
\mathcal{W}^\mathrm{H}(k_1,k_2)
= \frac{16\pi^2}{9} \left(\frac{\sqrt\pi}{\mu}\right)^3 k_1^3 k_2^3\,,
\end{equation}
and
\begin{eqnarray}
\mathcal{W}^\mathrm{F}(k_1,k_2)
&=& \frac{32\sqrt\pi^7}{3}
\left[ \mu\left\{ (k_1^2-k_1 k_2+k_2^2 - 2\mu^2)\,
e^{-\left(\frac{k_1+k_2}{2\mu}\right)^2}
\right.\right. \nonumber\\
&&\hspace*{2.5cm}\left.\left.- (k_1^2+k_1 k_2+k_2^2 - 2\mu^2)\,
e^{-\left(\frac{k_2-k_1}{2\mu}\right)^2}
\right\}\right. \nonumber\\
&&\hspace*{1.5cm}\left.- (k_1^3+k_2^3)
\,\mathrm{Erfc}\!\left(\frac{k_1+k_2}{2\mu}\!\right)
+ (k_2^3-k_1^3)
\,\mathrm{Erfc}\!\left(\frac{k_2-k_1}{2\mu}\!\right)
+ \sqrt\pi\,k_1^3\right]\,, \nonumber\\
\label{eq:W-G}\end{eqnarray}
where
\begin{equation} \mathrm{Erfc}(x) =
\int_x^\infty e^{-z^2} dz\,.
\end{equation}
In Eq.~(\ref{eq:W-G})
we have postulated $k_1\leq k_2$ again.
 \item \textit{Yukawa interaction}\\
For the Yukawa interaction we set
$f(r_{12})=e^{-\mu r_{12}}/\mu r_{12}$,
leading to $\tilde f(q)=4\pi/\mu(\mu^2+q^2)$.
This yields
\begin{equation}
\mathcal{W}^\mathrm{H}(k_1,k_2)
= \frac{64\pi^3}{9\mu^3} k_1^3 k_2^3\,,
\end{equation}
and
\begin{eqnarray}
\mathcal{W}^\mathrm{F}(k_1,k_2)
&=& \frac{2\pi^3}{3\mu} \left[ 4k_1 k_2
\left\{3(k_1^2+k_2^2)-\mu^2\right\}\right. \nonumber\\
&&\hspace*{1cm}\left. -16\mu\left\{ (k_1^3+k_2^3)
\arctan\!\left(\frac{k_1+k_2}{\mu}\!\right) - (k_2^3-k_1^3)
\arctan\!\left(\frac{k_2-k_1}{\mu}\!\right)
\right\} \right. \nonumber\\
&&\hspace*{1cm}\left. - \left\{3(k_2^2-k_1^2)^2
-6\mu^2(k_1^2+k_2^2)-\mu^4\right\}\,
\ln\frac{\mu^2+(k_1+k_2)^2}{\mu^2+(k_2-k_1)^2}\right]\,.
\end{eqnarray}
 \item \textit{Momentum-dependent $\delta$ interaction}\\
In the Skyrme interaction we have momentum-dependent terms
with the form $\frac{1}{2}\{\mathbf{p}_{12}^2\delta(\mathbf{r}_{12})
+\delta(\mathbf{r}_{12})\mathbf{p}_{12}^2\}$
and $\mathbf{p}_{12}\cdot\delta(\mathbf{r}_{12})\mathbf{p}_{12}$.
The former operates only on the even channels
and yields
\begin{equation}
\mathcal{W}^\mathrm{H}(k_1,k_2) = \mathcal{W}^\mathrm{F}(k_1,k_2)
= \frac{4\pi^2}{15} k_1^3 k_2^3 (k_1^2+k_2^2)\,.
\end{equation}
The latter acts on the odd channels, giving
\begin{equation}
\mathcal{W}^\mathrm{H}(k_1,k_2) = -\mathcal{W}^\mathrm{F}(k_1,k_2)
= \frac{4\pi^2}{15} k_1^3 k_2^3 (k_1^2+k_2^2)\,.
\end{equation}
\end{enumerate}

The incompressibility $\mathcal{K}$ and the spin-isospin curvatures
$a_t$, $a_s$, $a_{st}$ are
expressed by the derivatives of the $\mathcal{W}$ functions.

The single-particle energy $\epsilon(\mathbf{k}\sigma\tau)$
defined in Eq.~(\ref{eq:NM-spe}) is also expressed
by the derivative of the $\mathcal{W}$ functions.
We first rewrite the integral in Eq.~(\ref{eq:NME}) as
\begin{eqnarray}
 \int_{k'_1\leq k_1}d^3k'_1
  \int_{k_2\leq k_{\mathrm{F}\tau_2\sigma_2}}d^3k_2
  \langle\mathbf{k}'_1\sigma_1\tau_1,\mathbf{k}_2\sigma_2\tau_2
  |v_{12}|\mathbf{k}'_1\sigma_1\tau_1,\mathbf{k}_2\sigma_2\tau_2\rangle
\quad\quad\quad  \nonumber\\
  = 4\pi \int_0^{k_1}k'^2_1 dk'_1
  \int_{k_2\leq k_{\mathrm{F}\tau_2\sigma_2}}d^3k_2
  \langle\mathbf{k}'_1\sigma_1\tau_1,\mathbf{k}_2\sigma_2\tau_2
  |v_{12}|\mathbf{k}'_1\sigma_1\tau_1,\mathbf{k}_2\sigma_2\tau_2\rangle
  \,.
\end{eqnarray}
This immediately gives
\begin{eqnarray}
 \frac{\partial}{\partial k_1}
  \int_{k'_1\leq k_1}d^3k'_1
  \int_{k_2\leq k_{\mathrm{F}\tau_2\sigma_2}}d^3k_2
  \langle\mathbf{k}'_1\sigma_1\tau_1,\mathbf{k}_2\sigma_2\tau_2
  |v_{12}|\mathbf{k}'_1\sigma_1\tau_1,\mathbf{k}_2\sigma_2\tau_2\rangle
  \nonumber\\
 = 4\pi k_1^2 \int_{k_2\leq k_{\mathrm{F}\tau_2\sigma_2}}d^3k_2
  \langle\mathbf{k}'_1\sigma_1\tau_1,\mathbf{k}_2\sigma_2\tau_2
  |v_{12}|\mathbf{k}'_1\sigma_1\tau_1,\mathbf{k}_2\sigma_2\tau_2\rangle
  \,.\quad
\end{eqnarray}
Therefore,
\begin{eqnarray}
 \epsilon(\mathbf{k}_1\sigma_1\tau_1) &=&
  \frac{\mathbf{k}_1^2}{2M} + \frac{1}{(2\pi)^3} \frac{1}{4\pi k_1^2}
  \sum_n \sum_{\sigma_2\tau_2} \nonumber\\
&&\hspace*{2cm} \left[\left(t_n^{(\mathrm{W})}
+ t_n^{(\mathrm{B})} \delta_{\sigma_1\sigma_2}
- t_n^{(\mathrm{H})} \delta_{\tau_1\tau_2}
- t_n^{(\mathrm{M})} \delta_{\sigma_1\sigma_2} \delta_{\tau_1\tau_2}
\right) \partial_1\mathcal{W}_n^\mathrm{H}(k_1,
k_{\mathrm{F}\tau_2\sigma_2})
\right. \nonumber\\
&&\hspace*{2cm} \left. + \left(t_n^{(\mathrm{M})}
+ t_n^{(\mathrm{H})} \delta_{\sigma_1\sigma_2}
- t_n^{(\mathrm{B})} \delta_{\tau_1\tau_2}
- t_n^{(\mathrm{W})} \delta_{\sigma_1\sigma_2} \delta_{\tau_1\tau_2}
\right) \partial_1\mathcal{W}_n^\mathrm{F}(k_1,
k_{\mathrm{F}\tau_2\sigma_2}) \right]\,, \nonumber\\
\end{eqnarray}
where we use the short-hand notation
\begin{equation}
 \partial_1\mathcal{W}_n^\mathrm{H/F}(k_1,k_2)
  = \frac{\partial}{\partial k_1}\mathcal{W}_n^\mathrm{H/F}(k_1,k_2)\,.
\end{equation}
It is now obvious that the effective mass of Eq.~(\ref{eq:M*})
is expressed by using the second derivative
of the $\mathcal{W}$ functions.

\section{Landau parameters for symmetric nuclear matter\label{app:Landau}}

Let us denote the occupation probability of the s.p. states
of Eq.~(\ref{eq:NM-spwf}) by $n_{\tau\sigma}(\mathbf{k})$.
The nuclear matter energy of Eq.~(\ref{eq:NME-tot})
can be rewritten as
\begin{eqnarray}
\frac{\langle V\rangle}{\Omega}&=&
\frac{\langle V\rangle_\mathrm{H}+\langle V\rangle_\mathrm{F}}{\Omega}
\label{eq:pre-Lan}\\
&& \frac{\langle V\rangle_\mathrm{H}}{\Omega}
= \frac{1}{2(2\pi)^6} \sum_n \sum_{\sigma_1\sigma_2\tau_1\tau_2}
\sum_{\mathbf{k}_1\mathbf{k}_2} n_{\tau_1\sigma_1}(\mathbf{k}_1)
n_{\tau_2\sigma_2}(\mathbf{k}_2)\,\tilde f_n(0)
\nonumber\\
&&\hspace*{3cm} \cdot
\left(t_n^{(\mathrm{W})} + t_n^{(\mathrm{B})} \delta_{\sigma_1\sigma_2}
- t_n^{(\mathrm{H})} \delta_{\tau_1\tau_2}
- t_n^{(\mathrm{M})} \delta_{\sigma_1\sigma_2} \delta_{\tau_1\tau_2}
\right)\,,
\label{eq:pre-Lan-H}\\
&&\frac{\langle V\rangle_\mathrm{F}}{\Omega}
= \frac{1}{2(2\pi)^6} \sum_n \sum_{\sigma_1\sigma_2\tau_1\tau_2}
\sum_{\mathbf{k}_1\mathbf{k}_2} n_{\tau_1\sigma_1}(\mathbf{k}_1)
n_{\tau_2\sigma_2}(\mathbf{k}_2)\, \tilde f_n(2k_{12})
\nonumber\\
&&\hspace*{3cm} \cdot
\left(t_n^{(\mathrm{M})} + t_n^{(\mathrm{H})} \delta_{\sigma_1\sigma_2}
- t_n^{(\mathrm{B})} \delta_{\tau_1\tau_2}
- t_n^{(\mathrm{W})} \delta_{\sigma_1\sigma_2} \delta_{\tau_1\tau_2}
\right)\,.
\label{eq:pre-Lan-F}\end{eqnarray}
The Landau coefficient is defined by
\begin{equation} F_{\tau_1\sigma_1,\tau_2\sigma_2}^{(\ell)}(k_1,k_2)
= \frac{2\ell+1}{2} \int_{-1}^1 d(\hat{\mathbf{k}}_1\cdot\hat{\mathbf{k}}_2)
\,P_\ell(\hat{\mathbf{k}}_1\cdot\hat{\mathbf{k}}_2)\,
\frac{\delta^2(\langle V\rangle/\Omega)}
{\delta n_{\tau_1\sigma_1}(\mathbf{k}_1)
\delta n_{\tau_2\sigma_2}(\mathbf{k}_2)}\,.
\label{eq:def-Lan}\end{equation}
For the interaction independent of momentum and of density,
it is straightforward to write down
the coefficients of Eq.~(\ref{eq:def-Lan}) in terms of $\tilde f$,
within the HF theory at the zero temperature.
Noticing that $\rho$ also depends on $n_{\tau\sigma}(\mathbf{k})$,
we evaluate contribution of the density-dependent $\delta$ interaction
$(1+x^{(\mathrm{DD})} P_\sigma)\rho^\alpha \delta(\mathbf{r}_{12})$
to $F_{\tau_1\sigma_1,\tau_2\sigma_2}^{(\ell)}(k_1,k_2)$ as
\begin{eqnarray}
&& \frac{\delta_{\ell 0}}{(2\pi)^6} \left[
\frac{\alpha(\alpha-1)}{2} \rho^{\alpha-2}
\left\{\rho^2 - \sum_{\sigma\tau}\rho_{\tau\sigma}^2
+ x^{(\mathrm{DD})}\left(\sum_\sigma \rho_\sigma^2
		    - \sum_\tau \rho_\tau^2\right)
\right\} \right. \nonumber\\
&&\hspace*{1cm} \left. + \alpha\rho^{\alpha-1} \left\{
2\rho - \rho_{\tau_1\sigma_1} - \rho_{\tau_2\sigma_2}
+ x^{(\mathrm{DD})}(\rho_{\sigma_1}+\rho_{\sigma_2}
-\rho_{\tau_1}-\rho_{\tau_2})
\right\} \right. \nonumber\\
&&\hspace*{1cm} \left. + \rho^\alpha \left\{1
-\delta_{\tau_1\tau_2}\delta_{\sigma_1\sigma_2}
+ x^{(\mathrm{DD})}(\delta_{\sigma_1\sigma_2}
-\delta_{\tau_1\tau_2})\right\}
\right]\,,
\end{eqnarray}
where $\rho_\sigma=\sum_\tau \rho_{\tau\sigma}$
and $\rho_\tau=\sum_\sigma \rho_{\tau\sigma}$.
Apart from the spin and isospin degrees of freedom,
the momentum-dependent $\delta$ interactions
$\frac{1}{2}\{\mathbf{p}_{12}^2\delta(\mathbf{r}_{12})
+\delta(\mathbf{r}_{12})\mathbf{p}_{12}^2\}$
and $\mathbf{p}_{12}\cdot\delta(\mathbf{r}_{12})\mathbf{p}_{12}$
contribute to $F_{\tau_1\sigma_1,\tau_2\sigma_2}^{(\ell)}(k_1,k_2)$ by
\begin{equation} \frac{1}{(2\pi)^6}
\left(\delta_{\ell 0}\frac{k_1^2+k_2^2}{4}
- \delta_{\ell 1}\frac{k_1 k_2}{2}\right)\,.
\end{equation}

In characterizing effective interactions,
we view the Landau coefficients for the symmetric nuclear matter,
where $\rho_{\tau\sigma}=\rho/4$ for any $\tau$ and $\sigma$.
While formulae for the Landau parameters were derived
for the Skyrme interaction in Ref.~\cite{ref:Lan-Sky}
and for the Gogny interaction in Ref.~\cite{ref:Lan-Gog},
we here derive expressions for interactions
with the form of Eq.~(\ref{eq:effint}) in more general manner.
It is customary to transform the $(\tau,\sigma)$ variables
into the following ones,
\begin{equation}\begin{array}{ccl}
1 & \cdots & p\uparrow + p\downarrow + n\uparrow + n\downarrow\,, \\
t & \cdots & p\uparrow + p\downarrow - n\uparrow - n\downarrow\,, \\
s & \cdots & p\uparrow - p\downarrow + n\uparrow - n\downarrow\,, \\
st & \cdots & p\uparrow - p\downarrow - n\uparrow + n\downarrow\,.
\end{array}\end{equation}
Since $\sum_\sigma \sigma = \sum_\tau \tau = \sum_\sigma(\sigma\tau)
= \sum_\tau(\sigma\tau) = 0$,
all the off-diagonal coefficients with respect to $(1,t,s,st)$ vanish.
The diagonal coefficients are redefined as
\begin{eqnarray}
F_1^{(\ell)}(k_1,k_2) &=& \frac{1}{16}\sum_{\sigma_1\sigma_2\tau_1\tau_2}
F_{\tau_1\sigma_1,\tau_2\sigma_2}^{(\ell)}(k_1,k_2)\,,
\nonumber\\
F_t^{(\ell)}(k_1,k_2) &=& \frac{1}{16}\sum_{\sigma_1\sigma_2\tau_1\tau_2}
\tau_1\tau_2\,F_{\tau_1\sigma_1,\tau_2\sigma_2}^{(\ell)}(k_1,k_2)\,,
\nonumber\\
F_s^{(\ell)}(k_1,k_2) &=& \frac{1}{16}\sum_{\sigma_1\sigma_2\tau_1\tau_2}
\sigma_1\sigma_2\,F_{\tau_1\sigma_1,\tau_2\sigma_2}^{(\ell)}(k_1,k_2)\,,
\nonumber\\
F_{st}^{(\ell)}(k_1,k_2) &=& \frac{1}{16}\sum_{\sigma_1\sigma_2\tau_1\tau_2}
\sigma_1\tau_1\sigma_2\tau_2\,
F_{\tau_1\sigma_1,\tau_2\sigma_2}^{(\ell)}(k_1,k_2)\,.
\end{eqnarray}
The Hartree terms of the momentum- and density-independent interactions
yield
\begin{eqnarray}
F_{1,\mathrm{H}}^{(\ell)}(k_1,k_2) &=&
\frac{\delta_{\ell 0}}{4(2\pi)^6} \sum_n
(4t_n^{(\mathrm{W})}+2t_n^{(\mathrm{B})}
-2t_n^{(\mathrm{H})}-t_n^{(\mathrm{M})})
\,\tilde f_n(0)\,,
\nonumber\\
F_{t,\mathrm{H}}^{(\ell)}(k_1,k_2) &=&
\frac{\delta_{\ell 0}}{4(2\pi)^6} \sum_n
(-2t_n^{(\mathrm{H})}-t_n^{(\mathrm{M})})\,\tilde f_n(0)\,,
\nonumber\\
F_{s,\mathrm{H}}^{(\ell)}(k_1,k_2) &=&
\frac{\delta_{\ell 0}}{4(2\pi)^6} \sum_n
(2t_n^{(\mathrm{B})}-t_n^{(\mathrm{M})})\,\tilde f_n(0)\,,
\nonumber\\
F_{st,\mathrm{H}}^{(\ell)}(k_1,k_2) &=&
\frac{\delta_{\ell 0}}{4(2\pi)^6}
\sum_n (-t_n^{(\mathrm{M})})\,\tilde f_n(0)\,,
\end{eqnarray}
while the Fock terms
\begin{eqnarray}
F_{1,\mathrm{F}}^{(\ell)}(k_1,k_2) &=& \frac{1}{4(2\pi)^6} \sum_n
(4t_n^{(\mathrm{M})}+2t_n^{(\mathrm{H})}
-2t_n^{(\mathrm{B})}-t_n^{(\mathrm{W})})
\,G_n^{(\ell)}(k_1,k_2)\,,
\nonumber\\
F_{t,\mathrm{F}}^{(\ell)}(k_1,k_2) &=& \frac{1}{4(2\pi)^6} \sum_n
(-2t_n^{(\mathrm{B})}-t_n^{(\mathrm{W})})\,G_n^{(\ell)}(k_1,k_2)\,,
\nonumber\\
F_{s,\mathrm{F}}^{(\ell)}(k_1,k_2) &=& \frac{1}{4(2\pi)^6} \sum_n
(2t_n^{(\mathrm{H})}-t_n^{(\mathrm{W})})\,G_n^{(\ell)}(k_1,k_2)\,,
\nonumber\\
F_{st,\mathrm{F}}^{(\ell)}(k_1,k_2) &=& \frac{1}{4(2\pi)^6} \sum_n
(-t_n^{(\mathrm{W})})\,G_n^{(\ell)}(k_1,k_2)\,,
\label{eq:Fl-Fock}\end{eqnarray}
where
\begin{equation}
G_n^{(\ell)}(k_1,k_2) = \frac{2\ell+1}{2} \int_{-1}^1
d(\hat{\mathbf{k}}_1\cdot\hat{\mathbf{k}}_2)\,
P_\ell(\hat{\mathbf{k}}_1\cdot\hat{\mathbf{k}}_2)\,
\tilde f_n(2k_{12})\,.
\label{eq:G-fac}\end{equation}
Contribution of the density-dependent interaction
$t^{(\mathrm{DD})}(1+x^{(\mathrm{DD})} P_\sigma)\rho^\alpha
\delta(\mathbf{r}_{12})$ is given by
\begin{eqnarray}
F_{1,\mathrm{DD}}^{(\ell)}(k_1,k_2) &=& \frac{\delta_{\ell 0}}{4(2\pi)^6}
t^{(\mathrm{DD})}\frac{3(\alpha+1)(\alpha+2)}{2}\rho^\alpha\,,
\nonumber\\
F_{t,\mathrm{DD}}^{(\ell)}(k_1,k_2) &=& \frac{\delta_{\ell 0}}{4(2\pi)^6}
t^{(\mathrm{DD})}(-2x^{(\mathrm{DD})}-1)\rho^\alpha\,,
\nonumber\\
F_{s,\mathrm{DD}}^{(\ell)}(k_1,k_2) &=& \frac{\delta_{\ell 0}}{4(2\pi)^6}
t^{(\mathrm{DD})}(2x^{(\mathrm{DD})}-1)\rho^\alpha\,,
\nonumber\\
F_{st,\mathrm{DD}}^{(\ell)}(k_1,k_2) &=& -\frac{\delta_{\ell 0}}{4(2\pi)^6}
t^{(\mathrm{DD})}\rho^\alpha\,.
\end{eqnarray}
For momentum-independent interactions such as the Gogny interaction
and the M3Y-type interactions,
the Landau coefficients are obtained
by $F_1^{(\ell)}(k_1,k_2)=F_{1,\mathrm{H}}^{(\ell)}(k_1,k_2)
+F_{1,\mathrm{F}}^{(\ell)}(k_1,k_2)+F_{1,\mathrm{DD}}^{(\ell)}(k_1,k_2)$,
and so forth.
The momentum-dependent $\delta$ interactions yield
\begin{eqnarray}
F_{1,\mathrm{MD}}^{(\ell)}(k_1,k_2) &=& \frac{1}{8(2\pi)^6}
\left(\delta_{\ell 0}\frac{k_1^2+k_2^2}{2}-\delta_{\ell 1}k_1k_2\right)
\times\left\{\begin{array}{l}3t_1^{(\mathrm{MD})}\\5t_2^{(\mathrm{MD})}
\end{array}\right.\,,
\nonumber\\
F_{t,\mathrm{MD}}^{(\ell)}(k_1,k_2) &=& \frac{1}{8(2\pi)^6}
\left(\delta_{\ell 0}\frac{k_1^2+k_2^2}{2}-\delta_{\ell 1}k_1k_2\right)
\times\left\{\begin{array}{l}t_1^{(\mathrm{MD})}(-2x_1^{(\mathrm{MD})}-1)\\
t_2^{(\mathrm{MD})}(2x_2^{(\mathrm{MD})}+1)\end{array}\right.\,,
\nonumber\\
F_{s,\mathrm{MD}}^{(\ell)}(k_1,k_2) &=& \frac{1}{8(2\pi)^6}
\left(\delta_{\ell 0}\frac{k_1^2+k_2^2}{2}-\delta_{\ell 1}k_1k_2\right)
\times\left\{\begin{array}{l}t_1^{(\mathrm{MD})}(2x_1^{(\mathrm{MD})}-1)\\
t_2^{(\mathrm{MD})}(2x_2^{(\mathrm{MD})}+1)\end{array}\right.\,,
\nonumber\\
F_{st,\mathrm{MD}}^{(\ell)}(k_1,k_2) &=& \frac{1}{8(2\pi)^6}
\left(\delta_{\ell 0}\frac{k_1^2+k_2^2}{2}-\delta_{\ell 1}k_1k_2\right)
\times\left\{\begin{array}{l}(-t_1^{(\mathrm{MD})})\\t_2^{(\mathrm{MD})}
\end{array}\right.\,,
\label{eq:Lan-MD}\end{eqnarray}
where the upper row corresponds to the even channel interaction
$\frac{1}{2}t_1^{(\mathrm{MD})}(1+x_1^{(\mathrm{MD})} P_\sigma)
\{\mathbf{p}_{12}^2\delta(\mathbf{r}_{12})
+\delta(\mathbf{r}_{12})\mathbf{p}_{12}^2\}$,
while the lower to the odd channel interaction
$t_2^{(\mathrm{MD})}(1+x_2^{(\mathrm{MD})} P_\sigma)\mathbf{p}_{12}\cdot
\delta(\mathbf{r}_{12})\mathbf{p}_{12}$, respectively.
Equation~(\ref{eq:Lan-MD}) is available
for the Skyrme interactions in which the LS currents are not ignored.

We next show explicit form of the $G^{(\ell)}$ factor
in Eq.~(\ref{eq:Fl-Fock}) for typical interaction forms.
\begin{enumerate}
 \item \textit{$\delta$ interaction}\\
Substituting $\tilde f(2k_{12})$ by $1$,
we obtain
\begin{equation} G^{(\ell)}(k_1,k_2) = \delta_{\ell 0}\,.
\end{equation}
 \item \textit{Gauss interaction}\\
Because $\tilde f(q)=({\sqrt\pi}/\mu)^3\, e^{-(q/2\mu)^2}$,
Eq.~(\ref{eq:G-fac}) leads to
\begin{eqnarray} G^{(\ell)}(k_1,k_2)
&=& \frac{(2\ell+1)\sqrt\pi^3}{\mu k_1 k_2} \sum_{m=0}^\ell
\frac{(\ell+m)!}{m!(\ell-m)!} \left(\frac{\mu^2}{k_1 k_2}\right)^m
\nonumber\\
&&\hspace*{2cm} \cdot
\left\{(-)^m\,e^{-\left(\frac{k_1-k_2}{2\mu}\right)^2}
- (-)^\ell\,e^{-\left(\frac{k_1+k_2}{2\mu}\right)^2}\right\}\,.
\end{eqnarray}
For $\ell=0$ and $1$, we have
\begin{eqnarray}
G^{(0)}(k_1,k_2) &=& \frac{\sqrt\pi^3}{\mu k_1 k_2}
\left\{e^{-\left(\frac{k_1-k_2}{2\mu}\right)^2}
- e^{-\left(\frac{k_1+k_2}{2\mu}\right)^2}\right\}\,, \\
G^{(1)}(k_1,k_2)
&=& \frac{3\sqrt\pi^3}{\mu k_1 k_2}
\left\{\left(1-\frac{2\mu^2}{k_1k_2}\right)
\,e^{-\left(\frac{k_1-k_2}{2\mu}\right)^2}
+ \left(1+\frac{2\mu^2}{k_1k_2}\right)
\,e^{-\left(\frac{k_1+k_2}{2\mu}\right)^2}\right\}\,.
\end{eqnarray}
 \item \textit{Yukawa interaction}\\
For the Yukawa interaction we use
$\tilde f(q)=4\pi/\mu(\mu^2+q^2)$.
Inserting it into Eq.~(\ref{eq:G-fac}), we obtain for even $\ell$
\begin{eqnarray} G^{(\ell)}(k_1,k_2)
&=& \frac{2\pi(2\ell+1)}{\mu^3}
\sum_{m=0}^{\ell/2} \left(\frac{\mu^2}{2k_1k_2}\right)^{2m+1}
(-)^{\ell/2-m} \frac{(\ell+2m-1)!!}{(2m)!(\ell-2m)!} 
\nonumber\\
&&\hspace*{1cm}\cdot \left[ \left(1+\frac{k_1^2+k_2^2}{\mu^2}\right)^{2m}
\ln\frac{\mu^2+(k_1+k_2)^2}{\mu^2+(k_2-k_1)^2} \right.\nonumber\\
&&\hspace*{1cm} \left. - \sum_{p=0}^{2m-1} \frac{(-)^p}{2m-p}\,
\frac{(2m)!}{p!(2m-p)!} \left(1+\frac{k_1^2+k_2^2}{\mu^2}\right)^p
\right. \nonumber\\
&&\hspace*{2cm}\left. \cdot \left\{
\left(1+\frac{(k_1-k_2)^2}{\mu^2}\right)^{2m-p}
- \left(1+\frac{(k_1+k_2)^2}{\mu^2}\right)^{2m-p} \right\}
\right] \,. \nonumber\\
\end{eqnarray}
and for odd $\ell$
\begin{eqnarray} G^{(\ell)}(k_1,k_2)
&=& \frac{2\pi(2\ell+1)}{\mu^3}
\sum_{m=0}^{(\ell-1)/2} \left(\frac{\mu^2}{2k_1k_2}\right)^{2m+2}
(-)^{(\ell-1)/2-m} \frac{(\ell+2m)!!}{(2m+1)!(\ell-2m-1)!} 
\nonumber\\
&&\hspace*{1cm}\cdot \left[ \left(1+\frac{k_1^2+k_2^2}{\mu^2}\right)^{2m+1}
\ln\frac{\mu^2+(k_1+k_2)^2}{\mu^2+(k_2-k_1)^2} \right.\nonumber\\
&&\hspace*{1cm} \left. - \sum_{p=0}^{2m} \frac{(-)^{p+1}}{2m+1-p}\,
\frac{(2m+1)!}{p!(2m+1-p)!} \left(1+\frac{k_1^2+k_2^2}{\mu^2}\right)^p
\right. \nonumber\\
&&\hspace*{2cm}\left. \cdot \left\{
\left(1+\frac{(k_1-k_2)^2}{\mu^2}\right)^{2m+1-p}
- \left(1+\frac{(k_1+k_2)^2}{\mu^2}\right)^{2m+1-p} \right\}
\right] \,. \nonumber\\
\end{eqnarray}
For $\ell=0$ and $1$, we have
\begin{eqnarray} G^{(0)}(k_1,k_2)
&=& \frac{\pi}{\mu k_1k_2}
\,\ln\frac{\mu^2+(k_1+k_2)^2}{\mu^2+(k_2-k_1)^2} \,,\\
G^{(1)}(k_1,k_2)
&=& \frac{3\pi}{2\mu(k_1k_2)^2}
\left[ (\mu^2+k_1^2+k_2^2)
\ln\frac{\mu^2+(k_1+k_2)^2}{\mu^2+(k_2-k_1)^2} - 4k_1k_2 \right] \,.
\end{eqnarray}
\end{enumerate}

Setting $k_1=k_2=k_{\mathrm{F}0}$
and using the estimated level density at the Fermi momentum
$N_0 = (2\pi)^6\cdot 2k_{\mathrm{F}0}M_0^\ast/\pi^2$,
we define the usual Landau parameters
\begin{eqnarray}
f_\ell = N_0 F_1^{(\ell)}(k_{\mathrm{F}0},k_{\mathrm{F}0})\,,\quad
f'_\ell = N_0 F_t^{(\ell)}(k_{\mathrm{F}0},k_{\mathrm{F}0})\,,\nonumber\\
g_\ell = N_0 F_s^{(\ell)}(k_{\mathrm{F}0},k_{\mathrm{F}0})\,,\quad
g'_\ell = N_0 F_{st}^{(\ell)}(k_{\mathrm{F}0},k_{\mathrm{F}0})\,.
\label{eq:Landau}\end{eqnarray}

The second derivatives of $\mathcal{E}$ at the saturation point
are connected to the Landau parameters.
The following relations are verified,
\begin{eqnarray}
 \frac{M^\ast_0}{M} = 1+\frac{1}{3}f_1\,,\quad
 \mathcal{K} = \frac{3k_{\mathrm{F}0}^2}{M^\ast_0}(1+f_0)\,,\quad
 a_t = \frac{k_{\mathrm{F}0}^2}{6M^\ast_0}(1+f'_0)\,,\nonumber\\
 a_s = \frac{k_{\mathrm{F}0}^2}{6M^\ast_0}(1+g_0)\,,\quad
 a_{st} = \frac{k_{\mathrm{F}0}^2}{6M^\ast_0}(1+g'_0)\,.\quad\quad
\end{eqnarray}

\begin{acknowledgments}
This work is financially supported
as Grant-in-Aid for Scientific Research (C), No.~13640263,
by the Ministry of Education, Culture, Sports, Science and Technology,
Japan.
Numerical calculations are performed on HITAC SR8000
at Institute of Media and Information Technology, Chiba University,
at Information Technology Center, University of Tokyo,
and at Computing Center, Hokkaido University.
\end{acknowledgments}


\end{document}